\documentclass[onecolumn]{aastex61}

\usepackage{amsmath}
\renewcommand\vec{\boldsymbol}
\renewcommand\tensor{\boldsymbol}

\begin{document}

\title{Helium Energetic Neutral Atoms from the Heliosphere: Perspectives for Future Observations}
\correspondingauthor{Pawe\l{} Swaczyna}
\email{pswaczyna@cbk.waw.pl}

\author[0000-0002-9033-0809]{Pawe\l{} Swaczyna}
\affiliation{Space Research Centre of the Polish Academy of Sciences (CBK PAN), Bartycka 18A, 00-716 Warsaw, Poland}
\author{Stan Grzedzielski}
\affiliation{Space Research Centre of the Polish Academy of Sciences (CBK PAN), Bartycka 18A, 00-716 Warsaw, Poland}
\author{Maciej Bzowski}
\affiliation{Space Research Centre of the Polish Academy of Sciences (CBK PAN), Bartycka 18A, 00-716 Warsaw, Poland}

\begin{abstract}
Observations of energetic neutral atoms (ENAs) allow for remote sensing of plasma properties in distant regions of the heliosphere. So far, most of the observations have concerned only hydrogen atoms. In this paper, we present perspectives for observations of helium energetic neutral atoms (He ENAs). We calculated the expected intensities of He ENAs created by the neutralization of helium ions in the inner heliosheath and through the secondary ENA mechanism in the outer heliosheath. We found that the dominant source region for He ENAs is the inner heliosheath. The obtained magnitudes of intensity spectra suggest that He ENAs can be observed with future ENA detectors, as those planned on \emph{Interstellar Mapping and Acceleration Probe}. Observing He ENAs is most likely for energies from a few to a few tens of keV/nuc. Estimates of the expected count rates show that the ratio of helium to hydrogen atoms registered in the detectors can be as low as 1:$10^4$. Consequently, the detectors need to be equipped with an appropriate mass spectrometer capability, allowing for recognition of chemical elements. Due to the long mean free paths of helium ions in the inner heliosheath, He ENAs are produced also in the distant heliospheric tail. This implies that observations of He ENAs can resolve its structure, which seems challenging from observations of hydrogen ENAs since energetic protons are neutralized before they progress deeper in the heliospheric tail. 
\end{abstract}

\keywords{acceleration of particles --- ISM: abundances --- ISM: atoms --- solar wind --- space vehicles: instruments --- Sun: heliosphere }

\section{Introduction}\label{sec:introduction}
The solar wind emitted by the Sun creates a cavity in the local interstellar medium (LISM) called the heliosphere \citep{parker_1958, parker_1961}. The plasma in the heliosphere and its neighborhood is studied in situ using interstellar probes \emph{Voyager 1} and \emph{Voyager 2}. The measurements of the plasma performed along their trajectories are put in appropriate context by global observations of energetic neutral atoms (ENAs). The ENAs are created from energized ions in the plasma by charge-exchange processes with interstellar neutral atoms \citep{gruntman_1997}. Consequently, the ENAs allow for remote sensing of the heliospheric plasma. Such observations are available from \emph{Interstellar Boundary Explorer} \citep[\emph{IBEX},][]{mccomas_2009a} and the Ion and Neutral Camera (INCA) on board \emph{Cassini} \citep{krimigis_2009}.

\emph{IBEX} is equipped with two ENA detectors, one dedicated for lower energies \citep[\emph{IBEX}-Lo][]{fuselier_2009a} and the other for higher energies \citep[\emph{IBEX}-Hi][]{funsten_2009}. Since most of the ENAs are expected to be hydrogen atoms, higher mass elements are expected to be negligible in the observed flux of ENA. Consequently, \emph{IBEX}-Hi does not have a mass spectrometer capability. To analyze interstellar neutral atoms, \emph{IBEX}-Lo is able to recognize different species. Unfortunately, the fluxes of higher mass ENAs are not high enough to be distinguishable from hydrogen ENAs. Confirmed observations of the heliospheric helium energetic neutral atoms (He ENAs) are available only from HSTOF on board \emph{SOHO} \citep{hilchenbach_1998, czechowski_2012}. However, these observations are restricted to quiet times of the solar activity and are averaged over large swaths of the sky close to the ecliptic plane. 

In the perspective of the follow-up \emph{Interstellar Mapping and Acceleration Probe} (\emph{IMAP}) mission \citep{nrc_2013}, it is expedient to assess fluxes of higher mass ENAs. In this analysis, such an assessment is performed for He ENAs. Helium is the second most abundant chemical element in the solar wind and in the interstellar medium. Consequently, intensities of He ENAs should be high enough to be detectable.

The \emph{IBEX} observations show that the observed hydrogen ENAs originate from two mechanisms \citep{mccomas_2009, mccomas_2012a, mccomas_2014b}. The first source is the inner heliosheath, where pick-up ions (PUIs), created mostly in the supersonic solar wind and accelerated at the termination shock, are neutralized on the interstellar neutrals \citep{gruntman_1997}. The other mechanism is expected to be responsible for a bright structure extending over the sky called the \emph{IBEX} ribbon \citep{mccomas_2009}. This structure had not been expected prior to \emph{IBEX}. Among many potential sources proposed to explain the ribbon \citep[see review by][]{mccomas_2014a}, the secondary ENA mechanism seems the most likely \citep{schwadron_2009, heerikhuisen_2010}. In this mechanism, the observed ENAs originate mostly from the solar wind ions that are neutralized and create the primary ENA flux. The primary ENAs escape to the outer heliosheath, where they are ionized and re-neutralized. The re-neutralized atoms, called the secondary ENAs, can be directed in greater amounts backward to an observer at 1~au when the velocities of primary ENAs are almost perpendicular to the local interstellar magnetic field in the outer heliosheath. Consequently, the ribbon forms a nearly complete circular arc, which indicate the directions in which the interstellar magnetic field is perpendicular to the line of sights.

Estimations of the He ENA signal from the inner heliosheath were calculated by \citet{czechowski_2012} with the use of the hydrodynamical model by \citet{fahr_2000}. The obtained fluxes were compared with the HSTOF observation, but the agreement was not satisfactory. \citet{grzedzielski_2013} calculated the expected He ENA fluxes from the inner heliosheath in three models of the heliospheric plasma flow: the hydrodynamical model developed by \citet{izmodenov_2003}, the classical model by \citet{parker_1961}, and an ad hoc model obtained from scaling of the hydrodynamical model. In the follow-up analysis, \citet{grzedzielski_2014} showed that He ENAs observed by HSTOF can be explained using a simple model by \citet{suess_1990}, supplemented with a helium PUI spectrum fitted to agree with the observations from \emph{Voyagers}.

The method proposed by \citet{grzedzielski_2010a, grzedzielski_2013, grzedzielski_2014} treats helium ions in the inner heliosheath as test particles. The distribution of helium ions is split into multiple velocity bins and each of them is evolved separately along a flow line, with the energy-dependent cross sections included. This approach differs from that typically used for calculating the production of hydrogen ENAs in magnetohydrodynamic simulations, in which protons are split into several populations, each of these populations is given by a specified distribution, and only the temperatures and densities of the populations are allowed to vary in the inner heliosheath \citep[e.g.][]{heerikhuisen_2008, zank_2010, zirnstein_2014}. Nowadays, \citet{zirnstein_2015b} and \citet{zirnstein_2016b} used a similar approach to the one proposed by \citet{grzedzielski_2010a, grzedzielski_2013, grzedzielski_2014} and used in this analysis. 

The aim of this analysis is to provide estimations of intensities of He ENAs expected for observations at 1~au for a broad range of ENA energies from $\sim$0.5 keV/nuc to $\sim$100~keV/nuc. The adopted plasma flows are obtained from the scalar potentials proposed by \citet{suess_1990} and \citet{czechowski_1995}. Evolution of helium ions is calculated in the inner heliosheath, and the emerging He ENAs are created by neutralization processes. Additionally, the contribution of He ENAs created in the outer heliosheath through the secondary ENA mechanism is added. 

The methods of calculation of the He ENA signal are presented in Section~\ref{sec:ihs} for the inner heliosheath emission and in Section~\ref{sec:ohs} for the outer heliosheath emission. The resulting intensities are presented in Section~\ref{sec:results} and discussed in Section~\ref{sec:discussion}. The main conclusions from the analysis are presented in Section~\ref{sec:conclusions}. 

\section{He ENA from the inner heliosheath}\label{sec:ihs}
One of the two sources of He ENAs analyzed in this paper is the neutralization of helium ions on the neutral interstellar gas in the inner heliosheath. Here, the method used previously by \citet{grzedzielski_2010a, grzedzielski_2013, grzedzielski_2014} is developed for an extended energy range from 0.1 keV/nuc up to 1 MeV/nuc. 

The method of estimation of the He ENA signal consists of four steps. (1) Adoption of the appropriate model of the inner heliosheath, which describes the plasma density, velocity, and temperature organized by the solar wind flow lines. In addition, the model contains information about the distribution of the neutral interstellar hydrogen and helium in the heliosphere. (2) Assumption on spectra of helium ions just downstream of the termination shock. (3) Evolution of the spectra of helium ions along the flow lines. (4) Calculation of the ENA production from energetic ions neutralized on the interstellar neutral atoms and of their survival probabilities to reach 1~au. 

The analysis considers only a time-independent model and is based on the average conditions of the solar wind. The timescales of the addressed processes are of the order of the solar cycle. The results represent typical intensities expected for He ENAs. 

\subsection{Model of the plasma flow in the inner heliosheath}\label{sec:flow}
The plasma flow is modeled following the hydrodynamical description without a magnetic field, originally proposed by \citet{parker_1961}. He found that if the plasma flow can be treated as incompressible and irrotational, then its flow can be described by a potential $\Phi$, and the plasma bulk velocity can be obtained as a gradient of potential,\footnote{Some authors quoted in this section used a definition without the factor $1/\sqrt{n}$, here the form of their potential was uniformized.}
\begin{equation}
 \vec{u}=-\frac{1}{\sqrt{n}}\vec{\nabla}\Phi, \label{eq:pot2vel}
\end{equation}
where $n$ is the plasma density. While $n$ is originally the physical density, with the domination of protons, here it denotes the number density. \citet{parker_1961} chose the boundary conditions so that the interstellar medium far away from the Sun has a density $n_\mathrm{i}$ and moves in the $z$ direction with a bulk speed $u_\mathrm{i}$. The solar wind just downstream of the spherical termination shock, located at distance $r=r_\mathrm{s}$, has a density of $n_\mathrm{s}$ and a radial bulk speed $u_\mathrm{s}$. With these assumptions, the potential takes the form
\begin{equation}
 \Phi_\mathrm{P}=-\sqrt{n_\mathrm{i}}u_\mathrm{i}z+\sqrt{n_\mathrm{s}}u_\mathrm{s}\frac{r_\mathrm{s}^2}{r}. \label{eq:parkerpot}
\end{equation}
This potential was used by \citet{grzedzielski_2013} as one of three considered models of the plasma flow in the heliosphere. However, the model proposed by \citet{parker_1961} assumes that the distance to the termination shock should be much smaller than the distance to the stagnation point. This assumption is too strong for the heliosphere. \citet{suess_1990} proposed a generalization of this model to allow for any ratio of these distances. In their extension, the potential has the form
\begin{equation}
 \Phi_\mathrm{SN}=-\sqrt{n_\mathrm{i}}u_\mathrm{i}z\left[\frac{1}{2}\left(\frac{r_\mathrm{s}}{r}\right)^3+1 \right]+\sqrt{n_\mathrm{s}}u_\mathrm{s}\frac{r_\mathrm{s}^2}{r}. \label{eq:snpot}
\end{equation}
This potential differs only by the first term in the square bracket compared to the one given in Equation~\eqref{eq:parkerpot}. This potential was used by \citet{grzedzielski_2014}.

The above-mentioned potentials assume a purely hydrodynamical flow and do not include deceleration of the heliosheath plasma due to charge exchanges of plasma ions with the interstellar neutral atoms. \citet{czechowski_1995} proposed to add additional terms to the potentials $\Phi_\mathrm{P}$ or $\Phi_\mathrm{SN}$. If one neglects the mass loading contribution due to ionization by electron impact, which is probably low because of the low electron temperature (see the discussion below) but was still originally included by \citet{czechowski_1995}, then these terms take the form
\begin{equation}
 \Phi_\mathrm{CG}=\sqrt{n_\mathrm{s}}\left[\frac{1}{8}\frac{\alpha_\mathrm{cx}}{\gamma}+\frac{1}{2}b\right]r_\perp^2-\sqrt{n_\mathrm{s}}bz^2+\Phi_0, \label{eq:cgpot}
\end{equation}
where $r_\perp=\sqrt{r^2-z^2}$, $\Phi_0=\Phi_\mathrm{P}\text{ or }\Phi_\mathrm{SN}$, $b$ is an arbitrary parameter, and $\gamma=5/3$ is the adiabatic index. The charge-exchange rate $\alpha_\mathrm{cx}$ can be calculated as
\begin{equation}
 \alpha_\mathrm{cx}=n_\mathrm{H}\sigma_\mathrm{cx}v_\mathrm{rel}, \label{eq:cxrate}
\end{equation}
where $n_\mathrm{H}$ is the density of the neutral hydrogen in the inner heliosheath, $\sigma_\mathrm{cx}$ is the charge-exchange cross section, and $v_\mathrm{rel}$ is the relative speed of protons and hydrogen atoms. \citet{czechowski_1995} discussed the influence of the parameter $b$ on the flow: for $b>0$, the plasma bulk velocity for $z\to\-\infty$ diverges, for $b<0$, the tail is finite and ends in a singular point. They noticed that $b<0$ need not be excluded from consideration in the distant tail. Nonetheless, in this analysis it is assumed that $b=0$. \citet{zirnstein_2015b} and \citet{zirnstein_2016b} in their analyses calculated the flow lines effectively using the potential $\Phi_\mathrm{CG}$ with $\Phi_0=\Phi_\mathrm{P}$ and $b=-\alpha_\mathrm{cx}/(12\gamma)$, obtaining the additional term in a spherically symmetric form, but the flow has a sink point at a finite distance.

In this analysis, two flow potentials are considered. The first one ($\Phi_1$) is given by potential $\Phi_\mathrm{SN}$, in the other, ($\Phi_2$) terms connected to the charge exchange ($\Phi_\mathrm{CG}$) are added with $\Phi_0=\Phi_\mathrm{SN}$:
\begin{align}
 \Phi_1&=-\sqrt{n_\mathrm{i}}u_\mathrm{i}z\left[\frac{1}{2}\left(\frac{r_\mathrm{s}}{r}\right)^3+1 \right]+\sqrt{n_\mathrm{s}}u_\mathrm{s}\frac{r_\mathrm{s}^2}{r}, \label{eq:pot1}\\
 \Phi_2&=-\sqrt{n_\mathrm{i}}u_\mathrm{i}z\left[\frac{1}{2}\left(\frac{r_\mathrm{s}}{r}\right)^3+1 \right]+\sqrt{n_\mathrm{s}}u_\mathrm{s}\frac{r_\mathrm{s}^2}{r}+\sqrt{n_\mathrm{s}}\left[\frac{1}{8}\frac{\alpha_\mathrm{cx}}{\gamma}\right]r_\perp^2.\label{eq:pot2}
\end{align}
These models do not contain a magnetic field and have a rotational symmetry around the direction of motion of the interstellar medium. For both flows, the heliospheric termination shock is assumed to be spherical and located at $r_\mathrm{s}=90$~au from the Sun (possible effects of non-sphericity are discussed in	Section~\ref{sec:tail}). This is a midway value between the observed termination shock crossing by \emph{Voyager 2} at 83.7~au \citep{burlaga_2008, gurnett_2008} and by \emph{Voyager 1} at 94.0~au \citep{burlaga_2005, gurnett_2008}. The radial bulk speed downstream of the termination shock is set as $u_\mathrm{s}=150\ \mathrm{km\,s^{-1}}$, based on the \emph{Voyager 2} measurements after the termination shock crossing \citep{richardson_2011}. The density of the inner heliosheath plasma is adopted as $n_s=0.002\ \mathrm{cm^{-3}}$, i.e., close to the mean plasma density in the inner heliosheath measured so far by \emph{Voyager 2} \citep[][and later data from \url{ftp://spdf.gsfc.nasa.gov/pub/data/voyager/}]{richardson_2011}. The speed of the interstellar medium is set as $u_\mathrm{i}=-25.8\ \mathrm{km\,s^{-1}}$ \citep{bzowski_2015d}.

The potential $\Phi_2$ requires a value of the charge-exchange rate between proton and hydrogen atoms in the heliosheath. This rate is assumed constant throughout the heliosheath to facilitate the analytical approach. Most of the protons from the solar wind on the termination shock are moderately energized and have thermal speeds $\sim$30~$\mathrm{km\,s^{-1}}$, thus the relative velocity of the neutral hydrogen and the protons in the plasma is dominated by the difference of the bulk speeds. This relative velocity is set to $v_\mathrm{rel}=125\ \mathrm{km\,s^{-1}}$, which corresponds to the characteristic difference of the flows in the inner heliosheath. The cross section for the charge exchange calculated from the formula given by \citet{lindsay_2005} for an energy corresponding to this speed is $\sigma_\mathrm{cx}=3\times10^{-15}\ \mathrm{cm^{-2}}$. Together with the neutral hydrogen density in the inner heliosheath $n_\mathrm{H}=0.1\ \mathrm{cm}^{-3}$ \citep{bzowski_2009, zank_2013}, the charge-exchange rate is $\alpha_\mathrm{cx}=3.75\times10^{-9}\ \mathrm{s^{-1}}$. 

The considered models do not include the magnetic field, either in the heliosphere or in the interstellar medium. In consequence, the pressure balance on the heliopause need not reproduce the observations made by \emph{Voyager 1}. The interstellar plasma density $n_\mathrm{i}$ is tuned so that the distance to the heliopause in the \emph{Voyager 1} direction is reproduced rather than the actual value of this density obtained from heliospheric studies. The flows have rotational symmetry and the distances to the heliopause depend only on the angular distance from the nose, located at ecliptic $(\lambda_\mathrm{No},\,\beta_\mathrm{No})=(255\fdg8,\,5\fdg16)$. The \emph{Voyager 1} position at the heliopause in ecliptic coordinates was $(255\fdg5,\,35\fdg0)$, so the angular distance to the nose was $29\fdg8$. To reproduce the heliopause crossing at the distance of 121.6~au \citep{gurnett_2013}, one needs to adopt $n_\mathrm{i}=0.0659\ \mathrm{cm^{-3}}$ for the potential $\Phi_1$ and $n_\mathrm{i}=0.0641\ \mathrm{cm^{-3}}$ for the potential $\Phi_2$. The obtained values of $n_\mathrm{i}$ are in a relatively good agreement with the \emph{Voyager 1} observations of 0.08~$\mathrm{cm^{-3}}$ from the observed plasma oscillations \citep{gurnett_2013}. A summary of the adopted values of the flow potential parameters is presented in Table~\ref{tab:potparam}.

\begin{deluxetable*}{lrrrrrrrr}
  \tablecaption{Flow potential parameters\label{tab:potparam}}
  \tablehead{
    \colhead{} & 
    \colhead{$r_\mathrm{s}$} & 
    \colhead{$n_\mathrm{s}$} &
    \colhead{$u_\mathrm{s}$} &
    \colhead{$n_\mathrm{i}$} &
    \colhead{$u_\mathrm{i}$} &
    \colhead{$\alpha_\mathrm{cx}$} &
    \colhead{$b$} \\
    \colhead{} & 
    \colhead{(au)} & 
    \colhead{(cm$^{-3}$)} &
    \colhead{($\mathrm{km\,s^{-1}}$)} &
    \colhead{(cm$^{-3}$)} &
    \colhead{($\mathrm{km\,s^{-1}}$)} &
    \colhead{($10^{-9}$~s$^{-1}$)} &
    \colhead{($10^{-9}$~s$^{-1}$)}   
  }
  \startdata
  $\Phi_1$ 				& 90	& 0.002 & 150 & 0.0659 & -25.8 & 0    & 0\\
  $\Phi_2$ 				& 90	& 0.002 & 150 & 0.0641 & -25.8 & 3.75 & 0\\
  $\Phi_\mathrm{G2014}$	& 84.7	& 0.002 & 150 & 0.0700 & -25.0 & 0    & 0\\
  \enddata
  \tablecomments{Parameters for the flow potential given by Equation~\eqref{eq:cgpot} with $\Phi_0=\Phi_\mathrm{SN}$. $\Phi_\mathrm{G2014}$ present parameters used in \citet{grzedzielski_2014} for comparison.}
\end{deluxetable*}

\begin{figure}
  \epsscale{.6}
  \plotone{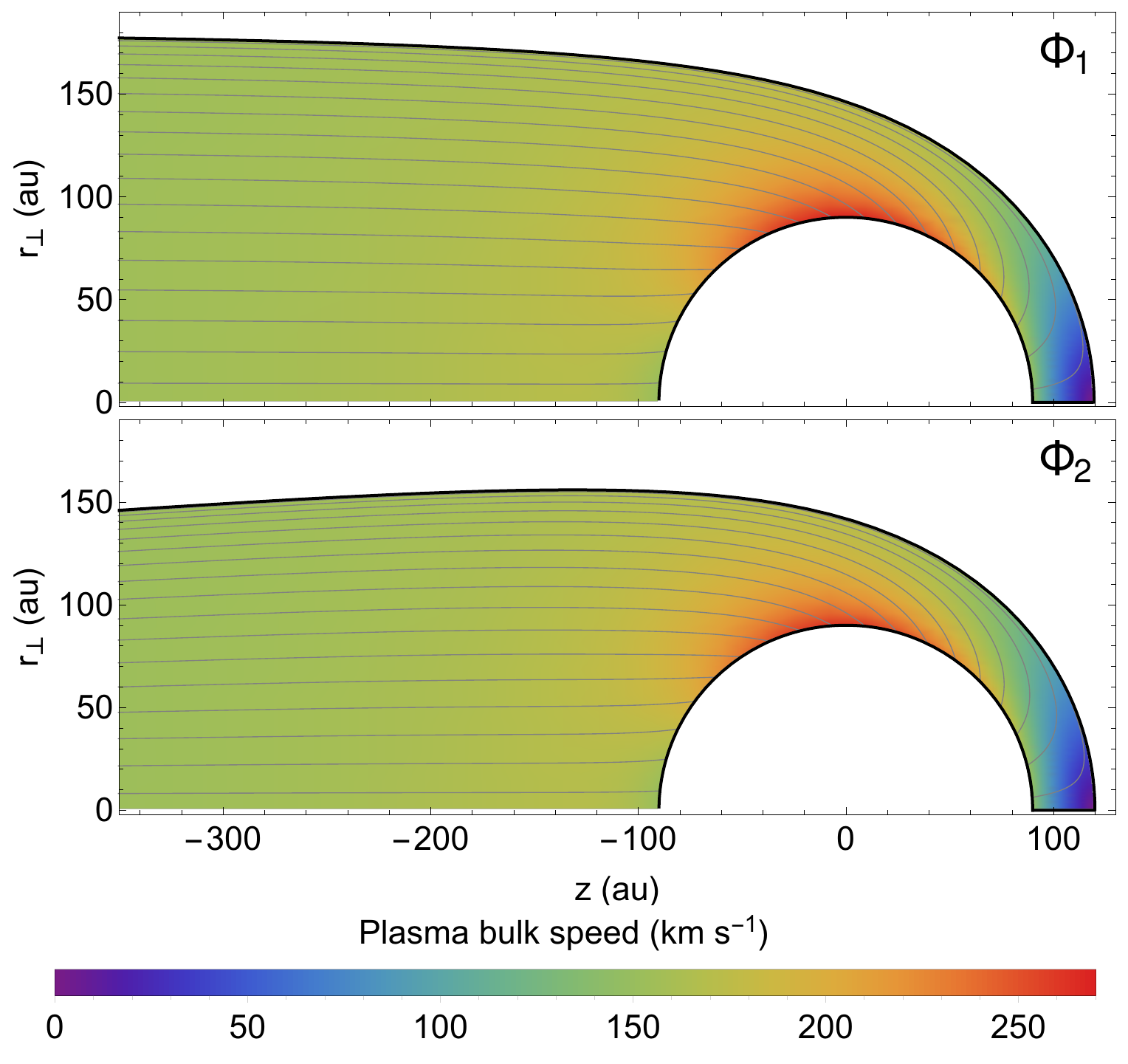}
  \caption{Plasma bulk speed in the inner heliosheath as given by potentials $\Phi_1$ (upper panel) and $\Phi_2$ (lower panel). The speed is the lowest in the frontal part of the heliosphere. The tail cross section is shrinking if the charge exchange loss is included.\label{fig:plasmaflow}}
\end{figure}

Figure~\ref{fig:plasmaflow} presents absolute values of the plasma bulk velocities in the inner heliosheath and the flow lines obtained from the two potentials. The flow obtained from the potential $\Phi_1$ has an almost constant width of the tail, whereas for $\Phi_2$ the tail is shrinking with the increasing distance from the Sun. This is a result of the plasma momentum loss due to the charge exchange included in the latter case.

\subsection{Plasma and neutrals in the inner heliosheath}
\label{sec:ihsdens}
The plasma flows obtained from the potentials do not determine the plasma parameters other than the velocity and density. In consequence, they need to be set separately. The solar wind plasma consists of protons, electrons, helium ions, and other heavy ions. The parameters of protons and electrons are discussed here and those of helium ions are discussed in subsequent sections. The contribution of other heavy ions is neglected in this analysis.

The electrons are assumed to follow the Maxwell--Boltzmann distribution with a thermal energy of 3 eV, based on the claim made by \citet{richardson_2011a} that the electron temperature in the inner heliosheath is below the 10 eV  threshold of the instrument on \emph{Voyager 2}. Most important is the fact that they have energies lower than that necessary for the electron impact ionization of the neutrals, which would inevitably cause extensive mass loading of protons to the plasma.

The protons are split into two populations: the solar wind protons and the PUIs. The solar wind protons are assumed to constitute 80\% of the total proton density and to have a temperature of 63500~K, which is adopted from the mean temperature observed by \emph{Voyager 1} \citep{richardson_2011}. The remaining 20\% are the PUIs, for which mean thermal energy of 1 keV is assumed. Both populations of protons in the inner heliosheath are assumed to follow the Maxwell--Boltzmann distribution. The PUIs have large charge-exchange rates and consequently they are depopulated along the flow lines. It is assumed that the density of the PUIs exponentially decreases along the flow line, so their density after time $t$ is $n_\mathrm{PUI}=n_\mathrm{PUI,TS}\exp(-\alpha_\mathrm{cx}t)$, where $\alpha_\mathrm{cx}=7.5\times10^{-9}\ \mathrm{s^{-1}}$ is calculated from Equation~\eqref{eq:cxrate} for 1 keV protons.

This charge-exchange process leads to replacement of the energetic PUIs with new ones, which have lower energies, resulting from the relative motion of the plasma and neutrals in the inner heliosheath. This energy is $\sim$0.1 keV, corresponding to the relative velocity of $\sim$125~$\mathrm{km\,s^{-1}}$. In the analysis of helium ENA production and losses, binary interactions with protons can be important only for higher energies (see Appendix~\ref{appendix:crosssec}). Consequently, the velocities of these $\sim$0.1 keV protons in the plasma frame are negligible compared to those of helium ions and atoms with which they interact. Eventually, these newly created PUIs are included in the core solar wind proton population. For the same reasons, the charge exchange between the solar wind protons and neutrals is neglected.

The interstellar neutral atoms are the most important source of the neutralization processes leading to the creation of ENAs. Here, it is assumed that neutral hydrogen and helium have a constant density throughout the inner heliosheath, equal $n_\mathrm{H}=0.1\ \mathrm{cm^{-3}}$ and $n_\mathrm{He}=0.015\ \mathrm{cm^{-3}}$ \citep{bzowski_2009, witte_2004, zank_2013}, respectively. The bulk speed of the interstellar neutral gas is also assumed to be constant and equal to the assumed speed of the plasma component in the interstellar medium $\vec{u}_\mathrm{H}=\vec{u}_\mathrm{He}=u_\mathrm{i}\vec{e}_z$. The temperature is set to $T_\mathrm{H}=T_\mathrm{He}=7440\ \mathrm{K}$ \citep{bzowski_2015d}. Note that especially for hydrogen, the secondary population created in the outer heliosheath is a significant portion of the overall neutral population. The parameters of the secondary population strongly differ from the given above \citep{izmodenov_2003}. However, the velocity and temperature of the secondary population would not significantly change the relative speeds of atoms and ions in the inner heliosheath used in the calculation of ENA production. 

\subsection{Helium ions spectra at the termination shock}\label{sec:heatts}
The populations of helium ions (He$^{2+}$ and He$^{+}$) downstream of the termination shock contain both helium ions emitted by the Sun and the helium PUIs created from the interstellar neutral helium by charge exchange with plasma ions or by photoionization. Section~\ref{sec:ssw} presents a model of the plasma and neutral populations upstream of the termination shock. Based on this model, the PUI fluxes at the termination shock are calculated (Section~\ref{sec:puiflux}). Finally, spectra of helium ions just downstream of the termination shock are developed (Section~\ref{sec:spectra}). 

\subsubsection{Plasma and neutrals upstream of the termination shock}\label{sec:ssw}
\begin{figure}
  \epsscale{.6}
  \plotone{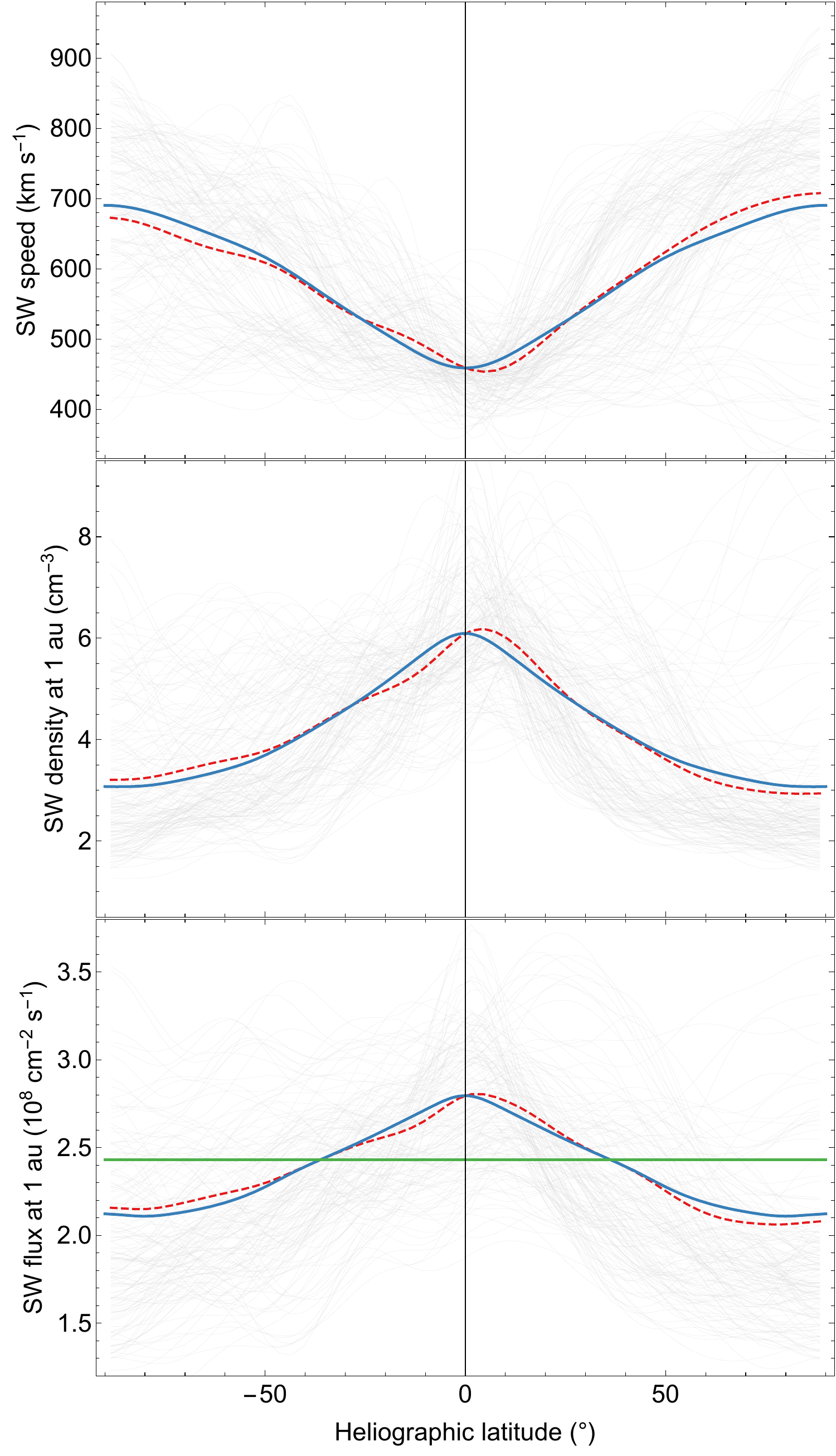}
  \caption{Profiles of the solar wind speed (upper panel), density (middle panel) and flux (lower panel) at 1 au, obtained from averaging of the profiles given by \citet{sokol_2015c} (profiles from solar cycle 23 are plotted as a bunch of light gray lines). The dashed red lines present profiles before symmetrization, and the solid blue lines are the symmetrized profiles. The flux of the solar wind is compared with the flux of the plasma downstream of the termination shock in the model presented in Section~\ref{sec:flow}, normalized to 1 au (green line).\label{fig:ssw}}
\end{figure}

In this analysis, the structure of the solar wind dependent on the heliographic latitude (hereafter denoted with $\theta$) is included, which affects the spectra of helium ions at the termination shock. Here, the model of the supersonic solar wind developed by \citet{sokol_2012, sokol_2015c} is adopted. Starting with this model, the latitudinal profiles of the solar wind speed and density are averaged over the period of the solar cycle 23 (i.e., between Carrington rotations 1909 and 2065). Subsequently, these profiles are symmetrized with respect to the solar equator, i.e., parameter values at latitude $\theta$ are obtained as means of their values for latitudes $\theta$ and $-\theta$. The obtained profiles are presented in Figure~\ref{fig:ssw}. The effect of symmetrization is smaller than 25~$\mathrm{km\, s^{-1}}$ for the speeds and 0.06~$\mathrm{cm^{-3}}$ for the density at 1 au. The resulting profiles of the density and bulk speed of the solar wind at 1 au are denoted as $n_\mathrm{SW,0}(\theta)$ and $u_\mathrm{SW,0}(\theta)$, respectively.

The flux of the plasma downstream of the termination shock in the model presented in Section~\ref{sec:flow} is equal to $n_\mathrm{s}u_\mathrm{s}=3\times10^4\ \mathrm{cm^{-2}s^{-1}}$. Normalized to the distance of 1 au, it is equivalent to a value of $n_\mathrm{s}u_\mathrm{s}r_\mathrm{s}^2/(1\ \mathrm{au})^2=2.43\times10^8\ \mathrm{cm^{-2}s^{-1}}$, which is close (within $\pm20\%$) to the mean flux obtained from the profiles of the supersonic solar wind. Although the model of the flow downstream of the termination shock does not account for the structure of the solar wind, the downstream flux is in agreement with the flux at 1 au. 

The solar wind is decelerated due to the generation of PUIs. This deceleration was predicted theoretically \citep{fahr_2001, fahr_2002} and observed by \emph{Voyager 2} \citep{richardson_2008}. Here, the bulk speed of the solar wind is assumed to decrease linearly with the distance $r$ according to the formula \citep{lee_2009}:
\begin{equation}
 u_\mathrm{SW}(r,\theta)=u_\mathrm{SW,0}(\theta)\left[1-\left(1-\frac{1}{2}\frac{\gamma-1}{2\gamma-1}\right)\frac{r}{\lambda_\mathrm{ml}} \right], \label{eq:vsw}
\end{equation}
where $\lambda_\mathrm{ml}$ is the characteristic length for mass loading, given by the formula: 
\begin{equation}
 \lambda_\mathrm{ml}=\left(\sigma_\mathrm{cx}^\mathrm{H^+,H}n_\mathrm{ISH}+\frac{\nu_\mathrm{H,0} n_\mathrm{ISH}+4\nu_\mathrm{He,0}n_\mathrm{ISHe}}{u_\mathrm{SW,0}(\theta)n_\mathrm{SW,0}(\theta)}\right)^{-1}. \label{eq:lml}
\end{equation}
For the charge-exchange cross section between protons and hydrogen atoms $\sigma_\mathrm{cx}^\mathrm{H^+,H}$ we use the formula given by \citet{lindsay_2005}\footnote{Cross sections depend on the relative velocities of the reagents, which is taken into account, but the dependency is not presented in formulas for brevity.}, $\nu_\mathrm{H,0}=1.1\times10^{-7}\ \mathrm{s^{-1}}$ and $\nu_\mathrm{He,0}=1.0\times10^{-7}\ \mathrm{s^{-1}}$ are the time-averaged photoionization rates at 1~au of hydrogen and helium atoms, respectively \citep{bzowski_2013b, bzowski_2015d, sokol_2014a}. In the calculation of this slowdown, constant values of the interstellar neutral hydrogen and helium in the heliosphere are adopted: $n_\mathrm{ISH}=0.09\ \mathrm{cm^{-3}}$ \citep{bzowski_2009} and $n_\mathrm{ISHe}=0.015\ \mathrm{cm^{-3}}$ \citep{witte_2004}. Based on this formula, the bulk speed decreases by $\sim$23\% up to the termination shock.

Additionally in the calculation of He$^{2+}$ PUI flux, the density of He$^{2+}$ ions emitted by the Sun is needed. The probability of neutralization of He$^{2+}$  is small, thus in the first order the flux is conserved, and consequently its density changes with distance as:
\begin{equation}
 n_\mathrm{SWHe^{2+}}(r,\theta)=0.05n_\mathrm{SW,0}(\theta)\frac{r_0^2v_\mathrm{SW,0}(\theta)}{r^2v_\mathrm{SW}(r,\theta)}, \label{eq:he2psw}
\end{equation}
where $r_0=1\ \mathrm{au}$, and it is assumed the that He$^{2+}$ ions constitute 5\% of the total plasma number density at 1 au \citep{kasper_2007}. 

The distribution of the interstellar neutral helium is strongly dependent on the position due to the formation of the focusing cone in the downwind direction. The focusing cone formed by the interstellar neutral helium population strongly affects the fluxes of the PUIs. Consequently, in the latter calculations we used a time-averaged distribution $n_\mathrm{ISHe}(r,\psi)$ of the neutral helium, which depends on the distance $r$ and the angular distance from the upwind direction $\psi$ \citep{sokol_2016}. 

\subsubsection{Flux of He PUIs}\label{sec:puiflux}
He$^+$ PUIs are produced mostly by photoionization, but a small contribution from the electron impact ionization is additionally included \citep{rucinski_1998}. The flux of He$^+$ PUIs at the termination shock is given by the integral:
\begin{equation}
 S_\mathrm{PUIHe^+}(\psi,\theta)=\frac{1}{r_\mathrm{s}^2}\int_0^{r_\mathrm{s}}n_\mathrm{ISHe}(r,\psi) \left[\nu_\mathrm{He,ph}(r)+\nu_\mathrm{He,el}(r,\theta)\right]r^2dr\, , \label{eq:fluxhe2}
\end{equation}
where $\theta$ is the heliographic latitude, $\nu_\mathrm{He,ph}(r)=\nu_\mathrm{He,0}(r_0/r)^2$ is the photoionization rate, and $\nu_\mathrm{He,el}(r,\theta)$ is the electron impact ionization rate. The electron impact ionization is important only close to the Sun due to the decrease of electron temperature with the distance from the Sun. We calculate this rate using the formula given by \citet{bzowski_2013b}.

He$^{2+}$ PUIs are created mostly by double charge exchange of the solar wind He$^{2+}$ ions with the neutral helium. This reaction has a negligible impact on the density of neutral helium, but is the main source of He$^{2+}$ PUIs. Their flux at the termination shock is equal to:
\begin{equation}
S_\mathrm{PUIHe^{2+}}(\psi,\theta)=\frac{1}{r_\mathrm{s}^2}\int_0^{r_\mathrm{s}}n_\mathrm{ISHe}(r,\psi)\sigma_\mathrm{2cx}^\mathrm{He^{2+},He}n_\mathrm{SWHe^{2+}}(r,\theta)v_\mathrm{rel}(\psi,\theta)r^2dr\, , \label{eq:fluxhe3}
\end{equation}
where $\sigma_\mathrm{2cx}^\mathrm{He^{2+},He}$ is the double charge-exchange rate between He$^{2+}$ ions and helium atoms (see Appendix~\ref{appendix:crosssec}), and $v_\mathrm{rel}(\psi,\theta)$ is the relative velocity between the solar wind plasma and the neutral helium, given by formula:
\begin{equation}
 v_\mathrm{rel}(r,\psi,\theta)=\left[ \left(u_\mathrm{SW}(r,\theta)\cos\psi-u_i\right)^2+\left(u_\mathrm{SW}(r,\theta)\sin\psi\right)^2\right]^{1/2}, \label{eq:relvel}
\end{equation}
where the thermal speed is neglected because it is small compared to the $u_\mathrm{SW}(r,\theta)$. 

Besides of the PUIs from the ionized interstellar neutral helium, the dominant component of helium ions is the solar wind contribution. Due to high temperature of the solar corona, the solar wind just emitted by the Sun contains mostly double ionized helium ions \citep{rucinski_1998}. The singly ionized component is produced by the neutralization of He$^{2+}$. The probability that a He$^{2+}$ ion is neutralized on its way to the termination shock is given by:
\begin{equation}
 p^\mathrm{He}_\mathrm{2\to1}(\psi,\theta)=1-\exp\left[-\int_0^{r_\mathrm{s}}\left( \sigma_\mathrm{cx}^\mathrm{He^{2+},He}n_\mathrm{ISHe}(r,\psi)+\sigma_\mathrm{cx}^\mathrm{He^{2+},H}n_\mathrm{ISH}\right)dr\right], \label{eq:probneutr}
\end{equation}
where $\sigma_\mathrm{cx}$ are cross sections for charge exchange with helium and hydrogen atoms (see Appendix~\ref{appendix:crosssec}). The charge exchange with hydrogen atoms is the dominant process for this neutralization. The probability varies from about 2.8\% to 9.2\% from the equator to the poles, respectively. This is a result of the dependence of the charge exchange on the relative velocity between ions and interstellar neutrals given by Equation~\eqref{eq:relvel}. Some of He$^{2+}$ originating from the Sun are completely neutralized by double charge exchange, and the respective probability is given by:
\begin{equation}
 p^\mathrm{He}_\mathrm{2\to0}(\psi,\theta)=1-\exp\left[-\int_0^{r_\mathrm{s}} \sigma_\mathrm{2cx}^\mathrm{He^{2+},He}n_\mathrm{ISHe}(r,\psi)dr\right]. \label{eq:probneutr2}
\end{equation}
This probability is lower and ranges from 0.4\% to 1.2\%. Based on these probabilities, the fluxes of helium ions at the termination shock are:
\begin{align}
 S_\mathrm{SWHe^{+},s}(\psi,\theta)&=0.05n_\mathrm{SW,0}(\theta)\frac{r_0^2u_\mathrm{SW,0}(\theta)}{r_\mathrm{s}^2}p^\mathrm{He}_\mathrm{2\to1}(\psi)\\
 S_\mathrm{SWHe^{2+},s}(\psi,\theta)&=0.05n_\mathrm{SW,0}(\theta)\frac{r_0^2u_\mathrm{SW,0}(\theta)}{r_\mathrm{s}^2}\left[1-p^\mathrm{He}_\mathrm{2\to1}(\psi)-p^\mathrm{He}_\mathrm{2\to0}(\psi)\right],
\end{align}
for He$^+$ and He$^{2+}$ ions respectively. 

The accumulated fluxes at the termination shock are related to the downstream density by the relation 
\begin{equation}
 n_{i,\mathrm{s}}(\psi,\theta)=S_i(\psi,\theta)/u_\mathrm{s}, \label{eq:puidens}
\end{equation}
where $u_\mathrm{s}$ is the radial component of the plasma bulk velocity downstream of the termination shock. Table~\ref{tab:heionsdens} presents densities of helium ion components downstream of the termination shock for some characteristic directions. The densities of He$^+$ and He$^{2+}$ ions from the solar wind change with heliographic latitude due to the large gradient of the cross section with the relative speeds (see Equation~\eqref{eq:relvel}). A more than twofold enhancement in the flux of He$^+$ PUIs is confined to the downwind region and is caused by the PUIs produced in the focusing cone of the interstellar neutral helium. In the upwind and crosswind directions (for $\psi<150\degr$) the obtained density is almost constant within $\pm6\%$. A similar asymmetry is visible for He$^{2+}$ PUI density, but here an additional factor is the cross section dependence on the relative velocity.

\begin{deluxetable*}{lrrrrrrrr}
  \tablecaption{Densities of helium ions downstream of the termination shock\label{tab:heionsdens}}
  \tablehead{
	\colhead{} & 
	\colhead{$\lambda$} & 
	\colhead{$\beta$} & 
	\colhead{$\psi$} & 
	\colhead{$\theta$} & 
	\colhead{$\mathrm{SW\ He^{+}}$} & 
	\colhead{$\mathrm{SW\ He^{2+}}$} & 
	\colhead{$\mathrm{PUI\ He^{+}}$} & 
	\colhead{$\mathrm{PUI\ He^{2+}}$}
	\\
	\colhead{} & 
	\colhead{($\degr$)} & 
	\colhead{($\degr$)} & 
	\colhead{($\degr$)} & 
	\colhead{($\degr$)} & 
	\colhead{($10^{-6}\ \mathrm{cm}^{-3}$)} & 
	\colhead{($10^{-6}\ \mathrm{cm}^{-3}$)} & 
	\colhead{($10^{-6}\ \mathrm{cm}^{-3}$)} & 
	\colhead{($10^{-6}\ \mathrm{cm}^{-3}$)} 
  }
  \startdata
  Upwind 			& 255.8	&   5.2 &   0  &   5.1 & 4.6 & 108.7 & 16.3 & 0.6 \\
  Downwind 			&  75.8	&  -5.2 & 180  &  -5.1 & 3.3 & 109.3 & 39.1 & 1.3 \\
  Sun's North Pole 	& 345.9 &  82.8 & 84.9 &  90   & 7.9 & 79.1 & 16.1 & 0.4 \\
  Crosswind 		& 165.1 &   7.2 &	90 &  0    & 3.8 & 110.6 &	16.2 &	0.6\\
  \emph{Voyager 1} 	& 255.0 &  35.0 & 29.8 &  34.6 & 6.6 & 93.7 & 16.2 & 0.5 \\
  \emph{Voyager 2} 	& 289.0 & -32.0 & 48.8 & -27.9 & 6.0 & 97.1 & 16.2 & 0.5 \\
  \enddata
  \tablecomments{The crosswind direction is one of the two directions perpendicular to the Sun's pole and to the upwind direction.}
\end{deluxetable*}

\subsubsection{Spectra of He ions downstream of the termination shock}\label{sec:spectra}
The energies of helium ions downstream of the termination shock are not known because \emph{Voyagers} were able to observe helium ions separately only for very high energies $\gtrsim 1\ \mathrm{MeV}$. Consequently, it is needed to determine the spectrum of helium ions based on some indirect arguments. In this analysis, it is assumed that the distribution functions of helium ions are isotropic in the plasma frame, i.e., the distribution function at a given position is a function of speed $v$ or equivalently, energy per nucleon $E$. 

The distribution of solar wind He$^+$ and He$^{2+}$ ions are assumed to follow the Maxwell--Boltzmann distribution with the mean energy per nucleon equal to that observed for the solar wind protons by \emph{Voyager~2} in the heliosheath. This corresponds to a temperature of 4$\times$63500~K (see Section~\ref{sec:flow}), or equivalently a mean energy of $\langle E_\mathrm{SW}\rangle=8.2\ \mathrm{eV/nuc}$. 

The PUIs at the termination shock are often described by two populations: transmitted PUIs and reflected PUIs. Reflection of a portion of PUIs is the dominating mechanism of energy dissipation at the termination shock \citep{zank_1996}. For protons, it is typically assumed that the energy at the termination shock is conserved and partitioned to three populations: the solar wind protons, the transmitted PUIs, and the reflected PUIs \citep{zank_2010, desai_2014}. \citet{zank_1996} showed that the fraction of the reflected PUIs scales as $\eta_\mathrm{r}\propto \sqrt{Z/M}$, where $Z$ is the charge and $M$ is the mass of an ion. For protons, the fraction of the reflected PUIs was calculated as $\sim$10\% of all hydrogen PUIs \citep{zank_2010}, and consequently the fraction of the reflected helium PUIs should be $\eta_\mathrm{r,He^+}=5\%$ and $\eta_\mathrm{r,He^{2+}}=7\%$ for He$^+$ and He$^{2+}$, respectively.

Helium ions are only a minority population in the solar wind. Consequently, conservation of their total energy is not a strict requirement for prediction of energies of helium ions downstream of the termination shock. Acceleration of ions at the termination shock is a complicated process \citep{zank_2010, zank_2015a}. Here, we assess energy of the transmitted PUIs assuming that they conserve their energies at the termination shock. Their thermal energies upstream of the termination shock are small compared to the bulk speed because of the adiabatic cooling in the expanding solar wind. Consequently, their energy at the termination shock is obtained from the difference of kinetic energy of the bulk flow in the supersonic solar wind and in the inner heliosheath:
\begin{equation}
 \langle E_\mathrm{PUI,t} \rangle(\theta)=\frac{m_\mathrm{p}}{2}\left[v_\mathrm{SW}^2(r_\mathrm{s},\theta)-v_\mathrm{s}^2\right], \label{eq:energypuitr}
\end{equation}
where $v_\mathrm{SW}(r_\mathrm{s},\theta)$ is the upstream speed, calculated from Equation~\eqref{eq:vsw}, and $v_\mathrm{s}$ is the downstream speed. The resulting energies of the transmitted PUIs range from $\sim$550~eV/nuc in the ecliptic plane to $\sim$1350~eV/nuc in the poles. The transmitted PUIs are assumed to follow the Maxwell--Boltzmann distribution, given by the formula:
\begin{equation}
 f_\mathrm{MB}(v_\mathrm{t};v)=\left(v_t^2\pi\right)^{-3/2}\exp\left(-\frac{v^2}{v_\mathrm{t}^2}\right), \label{eq:mbdist}
\end{equation}
where $v_\mathrm{t}$ is the characteristic speed related to the mean energy per nucleon in the plasma frame $\langle E\rangle$, given by the formula
\begin{equation}
v_t=\sqrt{\frac{4\langle E\rangle}{3m_\mathrm{p}}}. \label{eq:vt}
\end{equation}

The multiple reflections of the reflected PUIs are responsible for the creation of a high-energy tail of the PUI distribution \citep{zank_1996, giacalone_2010}. \citet{grzedzielski_2014} used the power-law spectrum of He$^+$ ions fit to the helium ion spectrum measured by \emph{Voyager 1} in the inner heliosheath in the first two months of 2007 \citep{stone_2008}. They obtained spectra of He ENAs in the four ecliptic sectors that are consistent with those observed by HSTOF \citep{czechowski_2012}. The HSTOF results covered the energy range 28--58~keV/nuc, and these authors concluded that the PUI spectrum at tens of keV/nuc can be estimated from the MeV observations of \emph{Voyager 1}. Here, we follow the same idea, but now we fit the $\kappa$-distribution given by the formula \citep{livadiotis_2013}
\begin{equation}
 f_\kappa(\kappa,v_\mathrm{t};v)=\frac{\Gamma(\kappa+1)}{\Gamma(\kappa-1/2)}\left[(\kappa-3/2)v_\mathrm{t}^2\pi\right]^{-3/2}\left(1+\frac{1}{\kappa-3/2}\frac{v^2}{v_\mathrm{t}^2}\right)^{-\kappa-1}, \label{eq:kapdist}
\end{equation}
where $\kappa$ is the index of the power-law tail of the intensity spectrum and $v_\mathrm{t}$ is the characteristic speed. Assuming that the \emph{Voyager 1} high-energy tail conforms with a $\kappa$-distribution with $\kappa=1.65$ \citep{decker_2005}, the density of PUIs in the \emph{Voyager 1} direction is as listed in Table~\ref{tab:heionsdens}, and the partition of the reflected PUIs is as mentioned above, one obtains that mean energy of $\langle E_\mathrm{PUI,r,V1} \rangle=19.1$~keV/nuc. This energy is scaled for directions different to the \emph{Voyager 1} direction with the energy of transmitted PUIs, as given by the formula
\begin{equation}
 \langle E_\mathrm{PUI,r} \rangle(\theta)=\langle E_\mathrm{PUI,r,V1} \rangle\frac{\langle E_\mathrm{PUI,t} \rangle(\theta)}{\langle E_\mathrm{PUI,t} \rangle(\theta_\mathrm{V1})},
\end{equation}
where $\theta_\mathrm{V1}=34.6^\circ$ is the heliographic latitude of \emph{Voyager 1}. The value of $\kappa$ for the reflected PUIs is assumed to be equal 1.65  everywhere at the termination shock. The mean energy for He$^+$ and He$^{2+}$ PUIs are assumed to be the same. 

\begin{figure}
  \epsscale{.6}
  \plotone{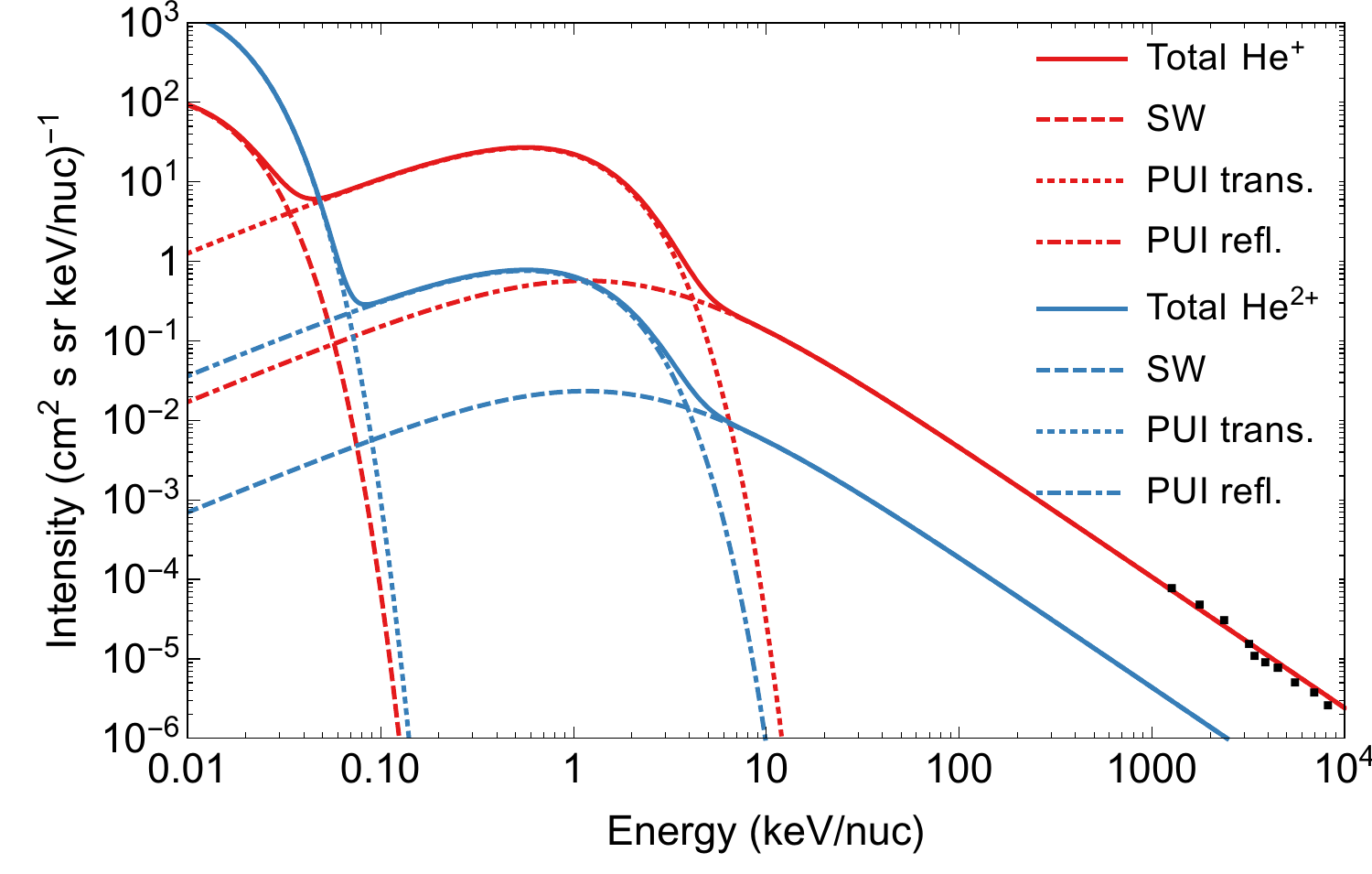}
  \caption{Intensity spectra of He$^+$ (red) and He$^{2+}$ (blue) ions in the \emph{Voyager 1} direction, evaluated from Equation~\eqref{eq:tsspectrum}. The spectra consist of the solar wind particles (dashed), the transmitted PUIs (dotted), and the reflected PUIs (dash-dotted).  The black squares in the bottom right corner are the \emph{Voyager 1} observations of helium ions, collected at time 2007/1-52 \citep{stone_2008}.\label{fig:tsspectrum}}
\end{figure}

The total spectrum of helium ions just downstream of the termination shock is a sum of the solar wind ions, the transmitted PUIs, and the reflected PUIs is
\begin{equation}
 f_i(\psi,\theta;v)=n_{\mathrm{SW}i,\mathrm{s}}(\psi,\theta)f_\mathrm{MB}( v_\mathrm{SW};v)+n_{\mathrm{PUI}i,\mathrm{s}}(\psi,\theta)\left[(1-\eta_{\mathrm{r,}i})f_\mathrm{MB}(v_\mathrm{PUI,t}(\theta);v)+\eta_{\mathrm{r,}i}f_\kappa(\kappa,v_\mathrm{PUI,r}(\theta);v)\right], \label{eq:tsspectrum}
\end{equation}
where $i=\mathrm{He^{+}\ or \ He^{2+}}$. The characteristic speeds $v_\mathrm{SW}$, $v_\mathrm{PUI,t}$, and $v_\mathrm{PUI,r}$ correspond to the mean energies $\langle E_\mathrm{SW}\rangle$, $\langle E_\mathrm{PUI,t}\rangle$, and $\langle E_\mathrm{PUI,r}\rangle$, respectively, as given by Equation~\eqref{eq:vt}. The intensity spectra of helium ions calculated at the \emph{Voyager 1} direction are presented in Figure~\ref{fig:tsspectrum}.

The total ion spectra weakly depend on the fraction of the reflected PUIs. This is caused by the low value of the $\kappa$ index assumed for the spectra of reflected PUIs, so these spectra have their maximum in the intensities for the energies $\sim$2~keV/nuc, where the dominating part comprises the transmitted PUIs. Adoption of various values for $\eta_{\mathrm{r,}i}\in(1\%,10\%)$ shows that the spectrum of the reflected PUIs changes only in the energy range where its contribution is at most comparable with the transmitted PUIs. The spectra of helium ions for energies 5--20~keV/nuc can have different shapes, which are not well described by a single $\kappa$-distribution. Unfortunately, the existing observations do not allow for verification of these spectra downstream of the termination shock.  

\subsection{Evolution of the spectra of helium ions along the flow lines}\label{sec:evolution}
The evolution of the distribution of helium ions $f_i=f_i(\vec{r},v)$ in the inner heliosheath is described by a transport equation adding the effects of spatial \citep{jokipii_1987} and velocity diffusion \citep{isenberg_1987} 
\begin{equation}
 \frac{\partial f_i}{\partial t}+\vec{u}\cdot\vec{\nabla}f_i=\frac{v}{3}\frac{\partial f_i}{\partial v}\left(\vec{\nabla}\cdot\vec{u}\right)+\frac{1}{v^2}\frac{\partial}{\partial v}\left( v^2D\frac{\partial f_i}{\partial v}\right)+\vec{\nabla}\cdot\left(\tensor{\kappa}\cdot\vec{\nabla}f_i\right)+\left.\frac{\delta f_i}{\delta t}\right|_\mathrm{c}, \label{eq:transport}
\end{equation}
where $\vec{u}$ is the flow velocity, $D$ is the velocity diffusion coefficient, and $\tensor{\kappa}$ is the tensor of spatial diffusion. In this analysis we consider stationary conditions, so the first term on the left side disappears. The second term describes advection with the plasma flow. The right-hand side terms describe: the adiabatic cooling/heating of the plasma, the diffusion in the velocity space, the spatial diffusion, and the collision loss/gain terms. 

The adiabatic term for the flow given by the potential $\Phi_1$ is zero, because $\vec{\nabla}\cdot \vec{u}=-n^{-1/2}\nabla^2\Phi_1=0$. The situation is different for the potential $\Phi_2$, where a nonzero divergence of the flow is caused by the charge exchange of the solar wind with the interstellar neutrals: $\vec{\nabla}\cdot \vec{u}=-n^{-1/2}\nabla^2\Phi_2=-\alpha_\mathrm{cx}/(2\gamma)$. 

The spatial diffusion is often neglected for the evolution of the PUI distribution in the inner heliosheath in similar approaches \citep{fahr_2004, zirnstein_2015b, zirnstein_2016b}. The spatial gradients of the distribution functions of helium ions at the termination shock are small because the distributions are obtained from the smooth formula given in Equation~\eqref{eq:tsspectrum}. The gradients along the flow lines are also limited because the charge-exchange processes, discussed further on, operate on longer scales (see Appendix~\ref{appendix:crosssec}). Consequently, the omission of the diffusion term is justified for helium ions. \citet{fahr_2004} argued that the phase space diffusion of the protons can be neglected. Details of such diffusion depend on the level of turbulence in the inner heliosheath and are out of scope of this analysis. 

The transport equation can be rewritten with the above assumption on the evolution of the distribution function along the flow line $f_i=f_i(s,v)$:
\begin{equation}
 u(s)\frac{\partial f_i}{\partial s}=\frac{v}{3}\frac{\partial f_i}{\partial v}\left(\vec{\nabla}\cdot\vec{u}\right)+\left.\frac{\delta f_i}{\delta t}\right|_\mathrm{c}, \label{eq:simtransport}
\end{equation}
where $u(s)$ is the flow speed along the flow line and $s$ is the length of the path along the flow line. The loss/gain terms describe binary interactions, which result in a change of the charge state of the ion. They can be expressed as
\begin{align}
 \left.\frac{\delta f_\mathrm{He^{+}}}{\delta t}\right|_\mathrm{c}&=
 -\alpha_\mathrm{1\to 0}f_\mathrm{He^{+}}
 -\alpha_\mathrm{1\to 2}f_\mathrm{He^{+}}
 +\alpha_\mathrm{2\to 1}f_\mathrm{He^{2+}}, \label{eq:collterm1}\\
 \left.\frac{\delta f_\mathrm{He^{2+}}}{\delta t}\right|_\mathrm{c}&=
 -\alpha_\mathrm{2\to 0}f_\mathrm{He^{2+}}
 -\alpha_\mathrm{2\to 1}f_\mathrm{He^{2+}}
 +\alpha_\mathrm{1\to 2}f_\mathrm{He^{+}}, \label{eq:collterm2}
\end{align}
where $\alpha_{i\to j}$ denotes the rates of processes that change the charge of a helium ion from $i$ to $j$. The first terms on the right-hand sides of Equations~\eqref{eq:collterm1} and~\eqref{eq:collterm2} represent the losses due to the ENA creation, the second terms are the losses of the population due to the change of the ion charge state $2\to1$ or $1\to2$, and the last terms are the gains to the other population from these processes. These rates depend on the relative speed of the reagents and the densities of the reagents at each position, and are described in detail in Appendix~\ref{appendix:crosssec}. The contribution of each process depends on the relative velocity of the reagents. Several different neutralization, charge exchange, and ionization processes are included in this analysis, but changes of the ion velocities due to charge exchange are neglected. Additionally, the contribution of the newly born PUIs in the inner heliosheath are also neglected, because their energies resulting from the relative velocity of the plasma and the interstellar neutral atoms are small and they do not contribute to the ENA fluxes in the energy range considered in this paper.

Equation~\eqref{eq:simtransport} is solved for $f_\mathrm{He^{+}}$ and $f_\mathrm{He^{2+}}$ along the flow lines obtained from both potentials separately, with the spectra assumed at the termination shock given by Equation~\eqref{eq:tsspectrum}. Details of the numerical scheme used are presented in Appendix~\ref{appendix:numeric}. 

\subsection{Integration of intensities of He ENAs}\label{sec:integrationheena}
Based on the distribution functions of helium ions in the inner heliosheath, $f_\mathrm{He^{+}}$ and $f_\mathrm{He^{2+}}$, the intensity of He ENAs in the Sun's frame at $r_0=1$~au can be calculated from the integral along the line of sight in the direction $\vec{\Omega}$, given by
\begin{equation}
 j_\mathrm{HeENA}^{\mathrm{IHS},v}(\vec{\Omega},v)=\int_{r_\mathrm{s}}^{r_\mathrm{hp}}w_\mathrm{sur}(r,v)v^2\left[f_\mathrm{He^{+}}\left(r\vec{\Omega},v'\right)\alpha_{1\to 0}+f_\mathrm{He^{2+}}\left(r\vec{\Omega},v'\right)\alpha_{2\to 0}\right] dr, \label{eq:enaintensityv}
\end{equation}
where $w_\mathrm{sur}(r,v)$ is the survival probability of He ENA on its path from the origin at $r$ to 1~au, $r\vec{\Omega}$ is the position in the inner heliosheath, and $v'$ is the speed of the parent ion in the plasma frame. The integral is calculated from the distance to the termination shock ($r_\mathrm{s}$) to the distance to the heliopause in the considered direction ($r_\mathrm{hp}$). The direction of observation $\vec{\Omega}$ and the velocities of He ENAs are approximately antiparallel because the ENAs are observed close to the Sun in comparison to the distance to their sources. Consequently, the speed of the parent ions in the plasma frame is given as
\begin{equation}
 v'=\left|-v\vec{\Omega}-\vec{u}\right|, \label{eq:vconversion}
\end{equation}
where $\vec{u}$ is the bulk velocity of the plasma flow. The survival probability is calculated from the equation:
\begin{equation}
 w_\mathrm{sur}(r,v)=\exp\left( -\int_{r_0}^{r} \frac{\alpha_{0\to2}+\alpha_{0\to1}}{v}dr'\right). \label{eq:survprob}
\end{equation}
The probabilities are calculated along the path of He ENAs from the source to $r_0=1$~au. 

The intensity given by Equation~\eqref{eq:enaintensityv} is a spectrum of He ENAs as a function of speed. The energy spectrum of He ENAs can be obtained from the transformation
\begin{equation}
 j^\mathrm{IHS}_\mathrm{HeENA}(\vec{\Omega},E)=j_\mathrm{HeENA}^{\mathrm{IHS},v}\left(\vec{\Omega},\sqrt{\frac{2E}{m_\mathrm{p}}}\right)\frac{1}{\sqrt{2m_\mathrm{p}E}}.  \label{eq:ihs}
\end{equation}

Details of the numerical scheme used to calculate the intensity of He ENAs are presented in Appendix~\ref{appendix:numeric}. \citet{grzedzielski_2013, grzedzielski_2014} gave a separate factor for the Compton--Getting effect at the source in the equations for the ENA intensity. This factor is effectively equal to $v^2/v'^2$ \citep{mccomas_2010}, which was included in Equation~\eqref{eq:enaintensityv}. Another Compton--Getting effect, due to the motion of the detector in the Sun's frame, is not discussed here. This can impact the ability of ENA detectors to measure these atoms. For the observational strategy similar to that used in \emph{IBEX}, half of the observations are made in the ram directions, where the number of observed atoms should be increased compared to that resulting from the inertial intensities (in the Sun's frame).  

\section{He ENA from the outer heliosheath}\label{sec:ohs}
The helium ions in the outer heliosheath can be a source of neutral atoms observed at 1 au. The thermal part of helium ions in the outer heliosheath after neutralization is a potential source of the secondary population of the interstellar neutral helium, discovered in the observations of the \emph{IBEX}-Lo detector as the Warm Breeze \citep{kubiak_2014, kubiak_2016, park_2016}.

The ribbon discovered by \emph{IBEX} in hydrogen ENAs \citep{mccomas_2009, fuselier_2009} should have a helium counterpart. \citet{swaczyna_2014} showed that expected intensities of the He ENAs created in the secondary ENA mechanism, which seems to be most likely an explanation for the IBEX ribbon, are comparable with the emission from the inner heliosheath. To calculate the expected He ENA signal created in this mechanism, a simple analytical model originally presented by \citet{mobius_2013} is employed, formulated by \citet{swaczyna_2016b}:
\begin{equation}
 j^\mathrm{OHS}_\mathrm{HeENA}(\vec{\Omega},E)=G(\vec{\Omega})I_\mathrm{SWHe,s}(\vec{\Omega},E)J(r_\mathrm{s},r_\mathrm{hp},\lambda_\mathrm{He},\tilde{\lambda}_\mathrm{He^{+}}), \label{eq:ohs}
\end{equation}
where $G(\vec{\Omega})$ is the geometric factor, $I_\mathrm{SWHe,s}(\vec{\Omega},E)$ is the flux of helium component of the neutral solar wind at the termination shock, and $J(r_\mathrm{s},r_\mathrm{hp},\lambda_\mathrm{He},\tilde{\lambda}_\mathrm{He^+})$ is the reflectance factor. \citet{swaczyna_2016b} assumed that the geometric factor has the form
\begin{equation}
 G(\vec{\Omega})=\frac{1}{2\pi\sigma_\mathrm{rib}}e^{-\frac{(\angle(\vec{\Omega},\vec{\Omega}_\mathrm{rib})-\phi_\mathrm{rib})^2}{2\sigma_\mathrm{rib}^2}}, \label{eq:geomfact}
\end{equation}
where $\angle(\vec{\Omega},\vec{\Omega}_\mathrm{rib})$ is the angular distance between the direction $\vec{\Omega}$ and the direction of the ribbon center in the ecliptic coordinates $\vec{\Omega}_\mathrm{rib}=(219\fdg2,39\fdg9)$, $\phi_\mathrm{rib}=74\fdg5$ is the mean ribbon cone angle, and $\sigma_\mathrm{rib}=10\fdg6$ is the mean width of the ribbon \citep{funsten_2013, schwadron_2014a}. The reflectance factor is defined as follows
\begin{equation}
 J(r_\mathrm{s},r_\mathrm{hp},\lambda_\mathrm{He},\tilde{\lambda}_\mathrm{He^+})=\frac{r_s^2}{\lambda_\mathrm{He}\tilde{\lambda}_\mathrm{He^+}}\int_{r_\mathrm{hp}}^\infty e^{-\frac{r_1-r_\mathrm{hp}}{\lambda_\mathrm{He}}}\left[ \int_{r_1}^\infty e^{-\frac{r_2-r_\mathrm{hp}}{\lambda_\mathrm{He}}}e^{-\frac{r_2-r_1}{\tilde{\lambda}_\mathrm{He^+}}} \frac{dr_2}{r_2^2}\right]dr_1. \label{eq:reflfact}
\end{equation}
The reflectance factor depends on the mean free path of helium atoms against ionization in the outer heliosheath,
\begin{equation}
 \lambda_\mathrm{He}=\frac{v}{\alpha_{0\to1}},
\end{equation}
and on the effective mean free path of He$^+$ ions against neutralization,
\begin{equation}
 \tilde{\lambda}_\mathrm{He^+}=\frac{|u_\mathrm{i}|\sin\theta_\mathrm{BV}}{\alpha_{1\to0}}, 
\end{equation}
where $\sin \theta_\mathrm{BV}$ is the angle between the inflow direction and the interstellar magnetic field, assumed here as $\theta_\mathrm{BV}=48\degr$. 

The differential flux of helium atoms in the neutral solar wind at the termination shock can be expressed as \citep{swaczyna_2016}
\begin{equation}
 I_\mathrm{SWHe,s}(\vec{\Omega},E)=\frac{1}{N}\sum_{i=1}^{N}\int_{r_0}^{r_\mathrm{s}}0.05n_{\mathrm{SW,0},i}(\theta)u_{\mathrm{SW,0},i}(\theta)\frac{r_0^2}{r_\mathrm{s}^2}\frac{e^{-r/\lambda_\mathrm{n}}}{\lambda_\mathrm{n}}\mathcal{N}\left(u_{\mathrm{SW},i}(r,\theta),\delta v|\sqrt{2E/m_\mathrm{p}} \right)\frac{1}{\sqrt{2m_\mathrm{p}E}}dr. \label{eq:isw}
\end{equation}
The above formula is adjusted to obtain the flux of neutral helium atoms resulting from neutralization of the supersonic solar wind. The index $i$ enumerates Carrington rotations from 1909 to 2065 over which the parameters of the solar wind are averaged. The original formula is multiplied by the factor 0.05 for the adopted abundance of He$^{2+}$ ions in the solar wind. The distribution of solar wind speeds is assumed to be given by the normal distribution $\mathcal{N}$ with standard deviation $\delta v=100\,\mathrm{km\,s^{-1}}$. The mean free path for neutralization of solar He$^{2+}$ ions is given by
\begin{equation}
\lambda_\mathrm{n}=\left(\sigma_\mathrm{2cx}^\mathrm{He^{2+},He}n_\mathrm{ISHe}(r,\psi)\right)^{-1}. \label{eq:lambdan}
\end{equation}
Contribution of the neutral solar wind helium resulting from two subsequent single neutralization processes is neglected. They contribute less than 10\% of the total flux of helium in the neutral solar wind at the termination shock \citep{swaczyna_2014}. Although the primary ENAs are produced both in the supersonic solar wind and in the inner heliosheath, here only the first source is included. The neutralized supersonic solar wind comprises enough primary ENAs to reproduce the ribbon observation in hydrogen \citep{swaczyna_2016b}. Contribution to the ribbon signal from the primary ENAs created in the inner heliosheath is not sufficient to reproduce the observations \citep{zirnstein_2016}.

\citet{swaczyna_2014} showed that the fluxes of He ENAs from the \emph{IBEX} ribbon in the secondary ENA mechanism are $\sim$10$^{-4}$ times smaller than the fluxes of hydrogen ENAs observed by \emph{IBEX} \citep{schwadron_2014a} and obtained from the analytic model similar to that used in this paper \citep{swaczyna_2016b}.

\section{Results}\label{sec:results}
\begin{figure*}
  \epsscale{1.15}
  
  \plotone{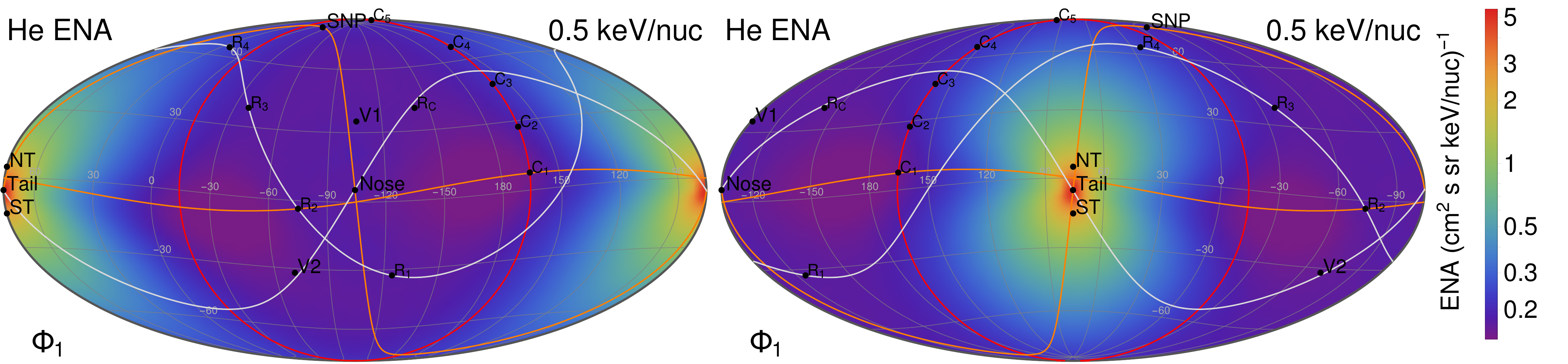}
 
  \plotone{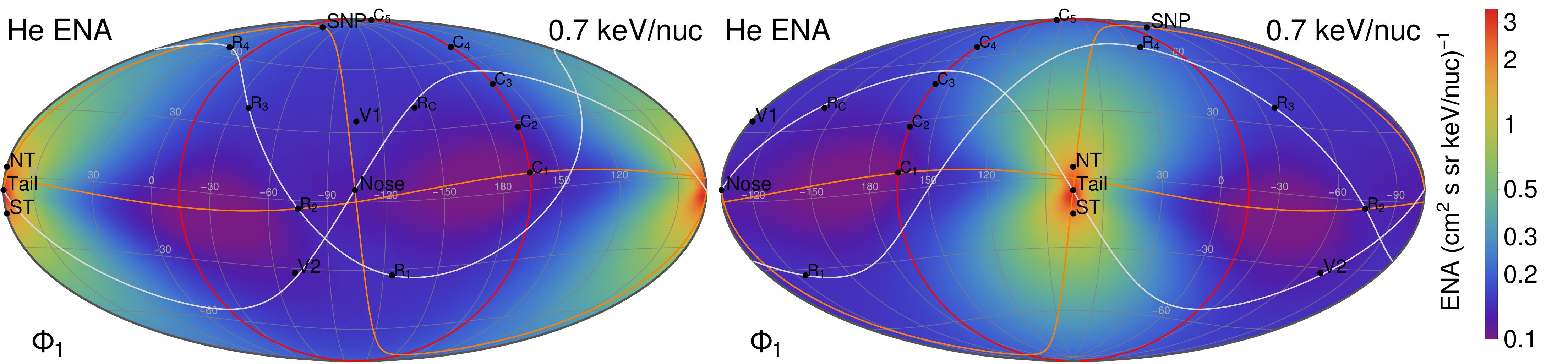}

  \plotone{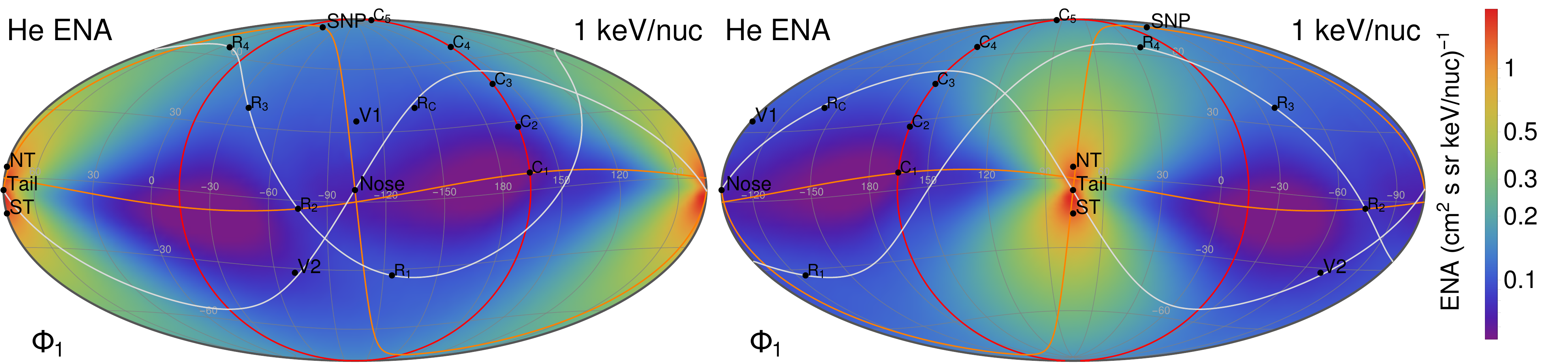}

  \plotone{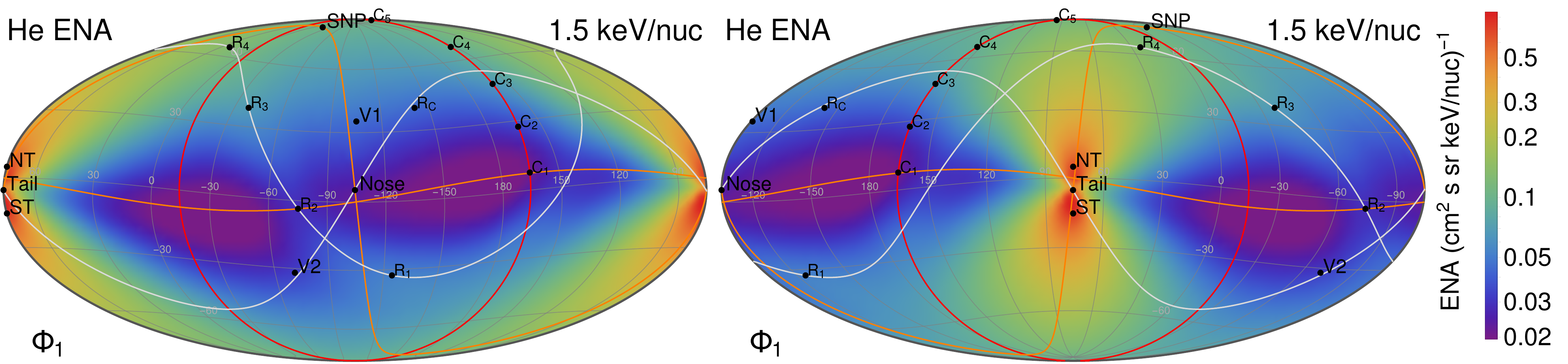}

  \plotone{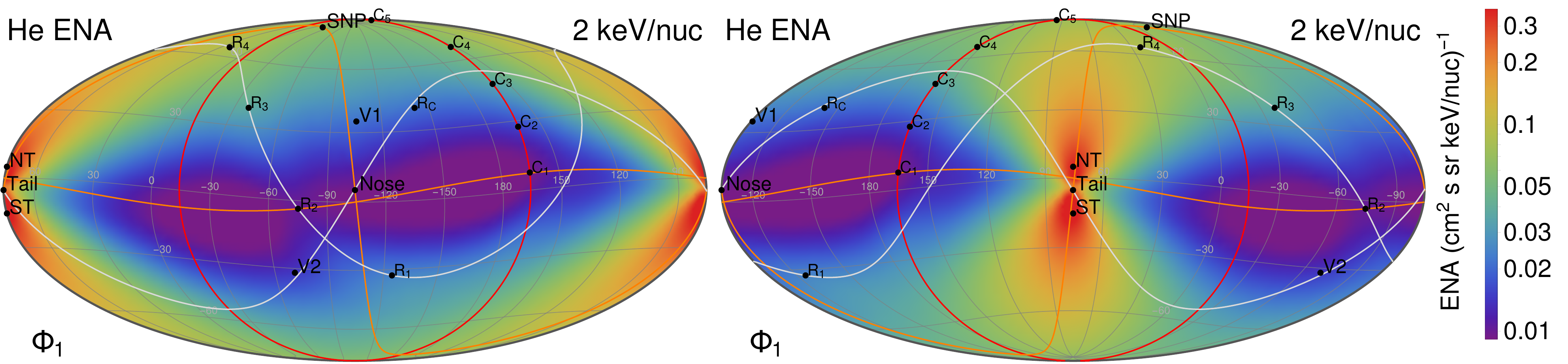}
  
  \caption{Maps of the expected differential intensities of He ENAs centered on the upwind direction (left column) and on the downwind direction (right column) for energies 0.5, 0.7, 1.0, 1.5, and 2.0 keV/nuc (top to bottom). The solid lines denote the ribbon circle and the deflection plane (both light gray), the crosswind directions (red), the solar equator, and the great circle containing the solar north pole and the upwind direction (both orange). $\mathrm{R}_1-\mathrm{R}_4$ and $\mathrm{C}_1-\mathrm{C}_5$ are selected directions, discussed in the text; the spectra for these directions are presented in Figure~\ref{fig:spectraheena}. \label{fig:mapsheena1}}
\end{figure*}

\begin{figure*}
  \epsscale{1.15}
  
  \plotone{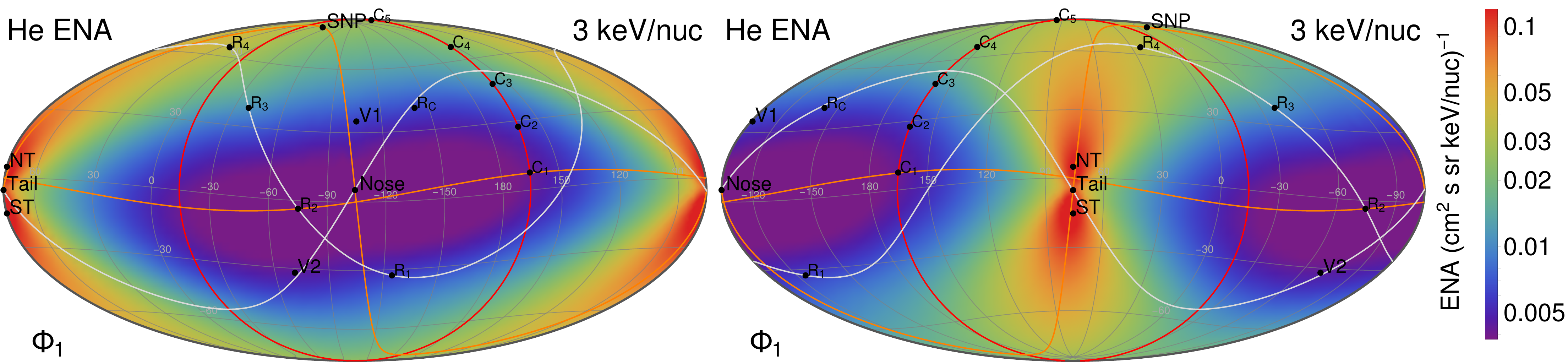}

  \plotone{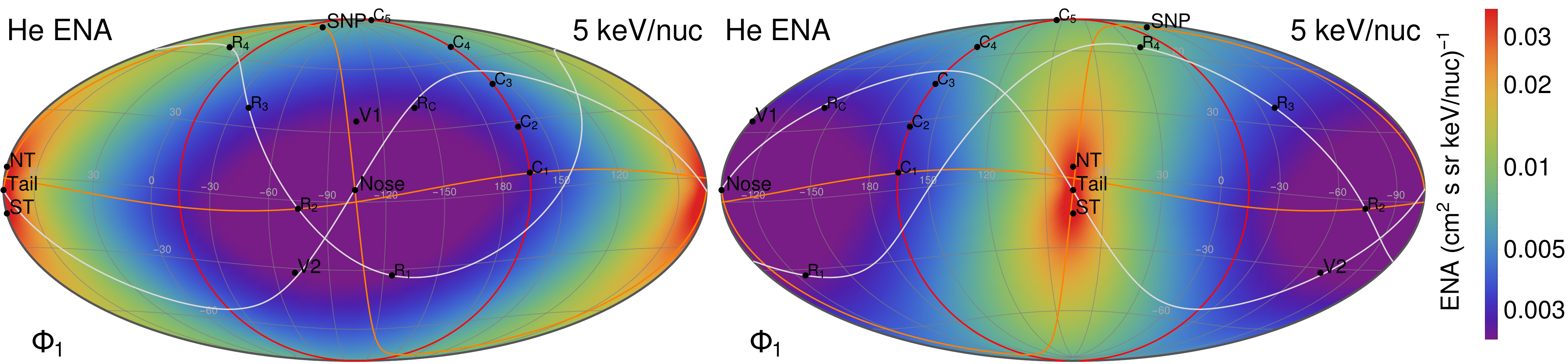}

  \plotone{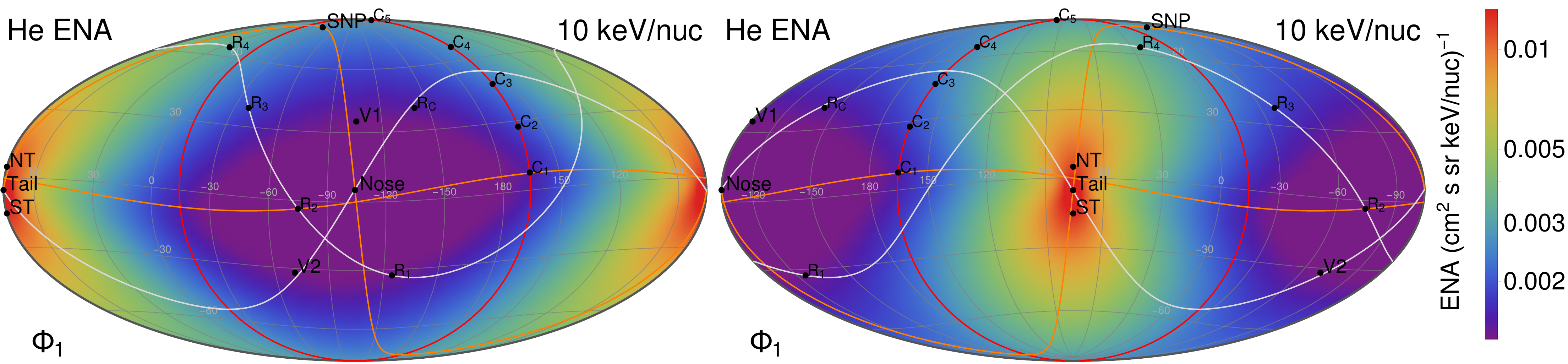}

  \plotone{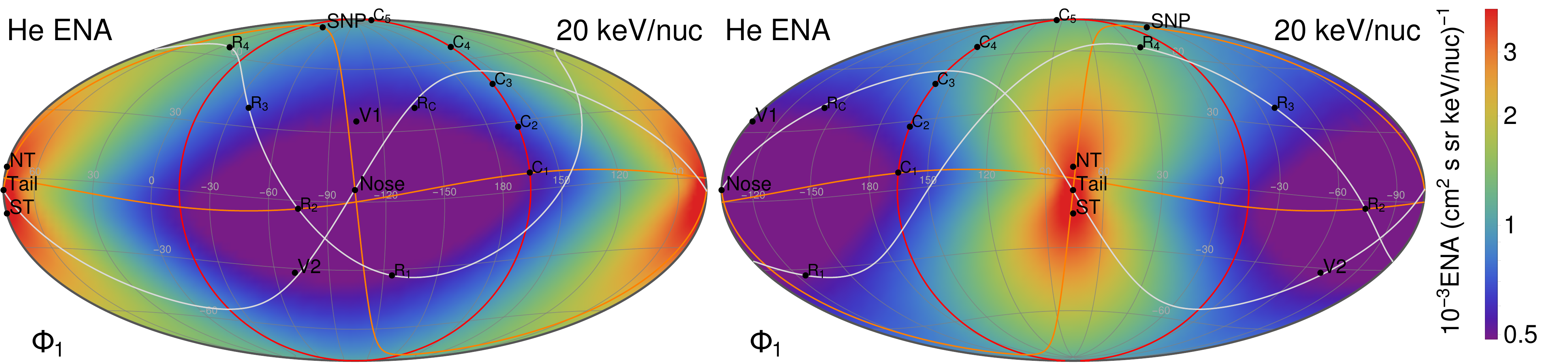}

  \plotone{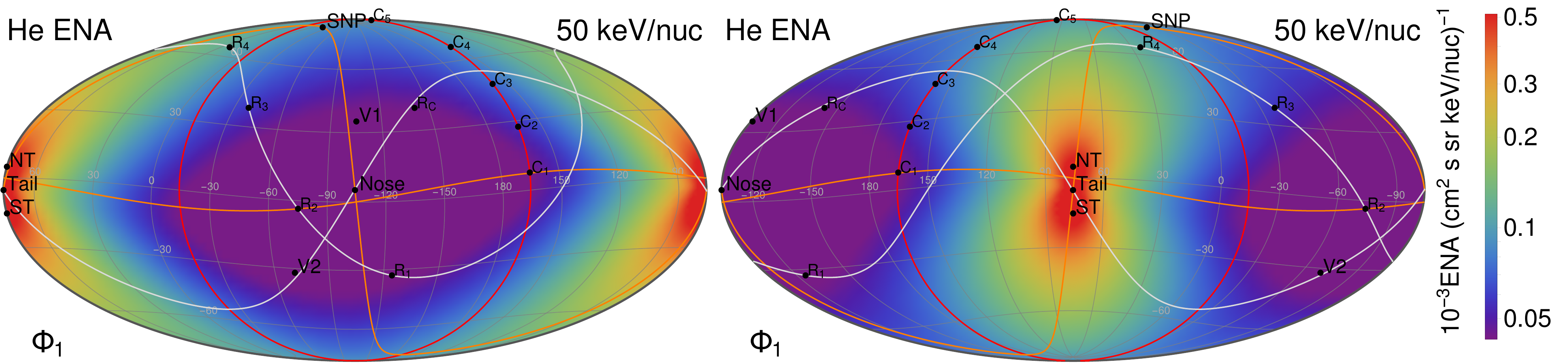}
  
  \caption{As in Figure~\ref{fig:mapsheena1}, but for energies 3, 5, 10, 20, and 50 keV/nuc (top to bottom).\label{fig:mapsheena2}}
\end{figure*}

The expected signal of He ENAs is calculated for the two plasma flows, obtained from the potentials $\Phi_1$ and $\Phi_2$ using Equation~\eqref{eq:pot2vel}. Sums of both components (the inner heliosheath and the ribbon) for the flow given by potential $\Phi_1$ are presented as the maps of intensities of He ENAs for the energies from 0.5 keV/nuc to 50 keV/nuc in Figures~\ref{fig:mapsheena1} and~\ref{fig:mapsheena2}. The maps are plotted in the Mollweide projection centered in the upwind (Nose) and downwind (Tail) directions of the Sun's motion respective to the interstellar medium. Due to the large span of intensities in each map the color scheme is logarithmic. 

The maps reveal a large tail-to-nose ratio of the intensities for all considered energies. The tailward part of the signal has a characteristic hourglass-shaped form, aligned with the meridian crossing the downwind direction. This effect is a result of the latitudinal structure of the PUIs downstream of the termination shock, as described in Section~\ref{sec:puiflux}. The signal for larger latitudes is larger than in the equatorial plane due to the contribution of the PUIs from the flow lines originating at higher heliographic latitudes and bent to the downwind direction. This effect is discussed further in Section~\ref{sec:hourglass}. 

\begin{figure}
  \epsscale{.6}
  
  \plotone{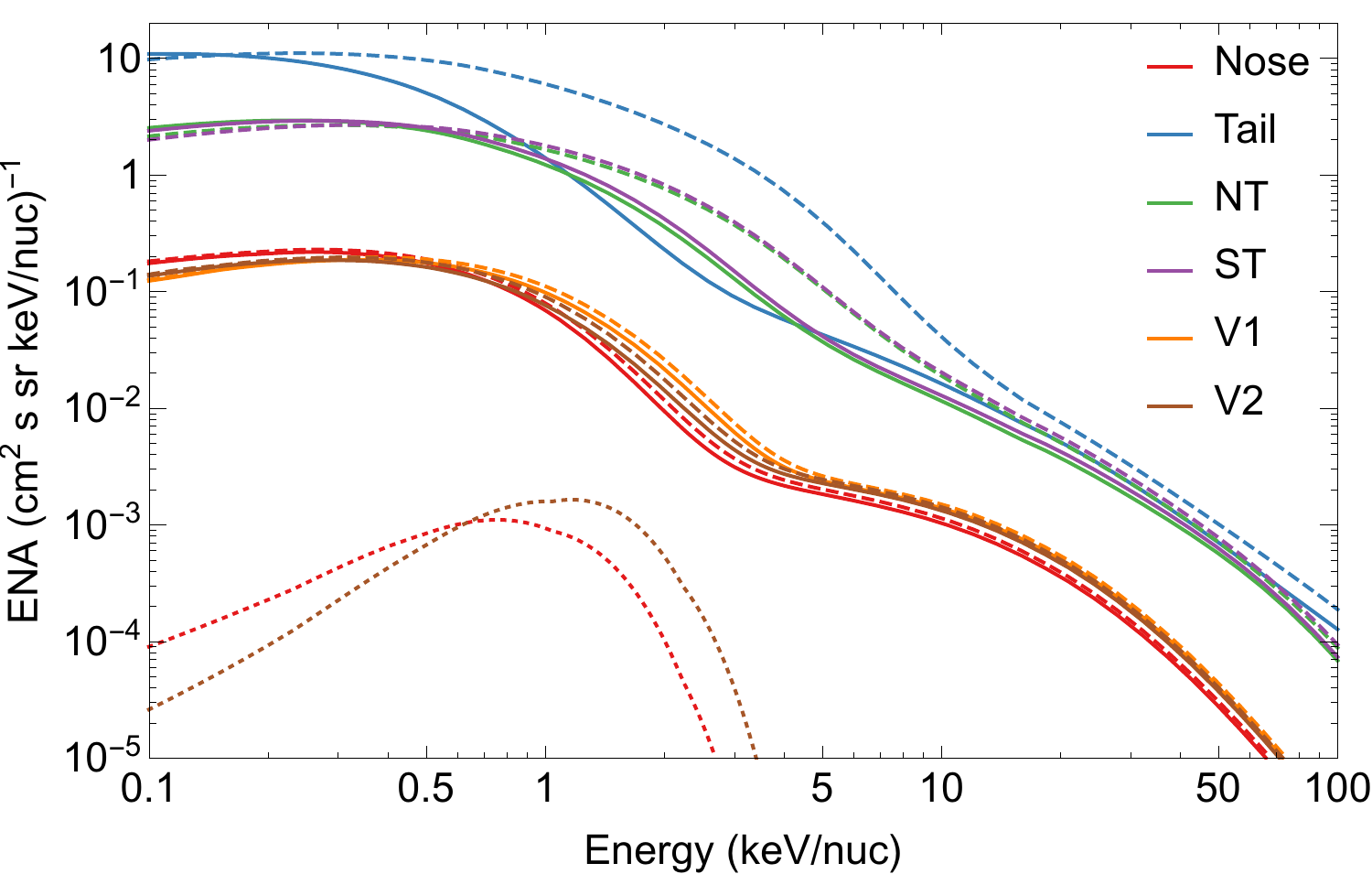}

  \plotone{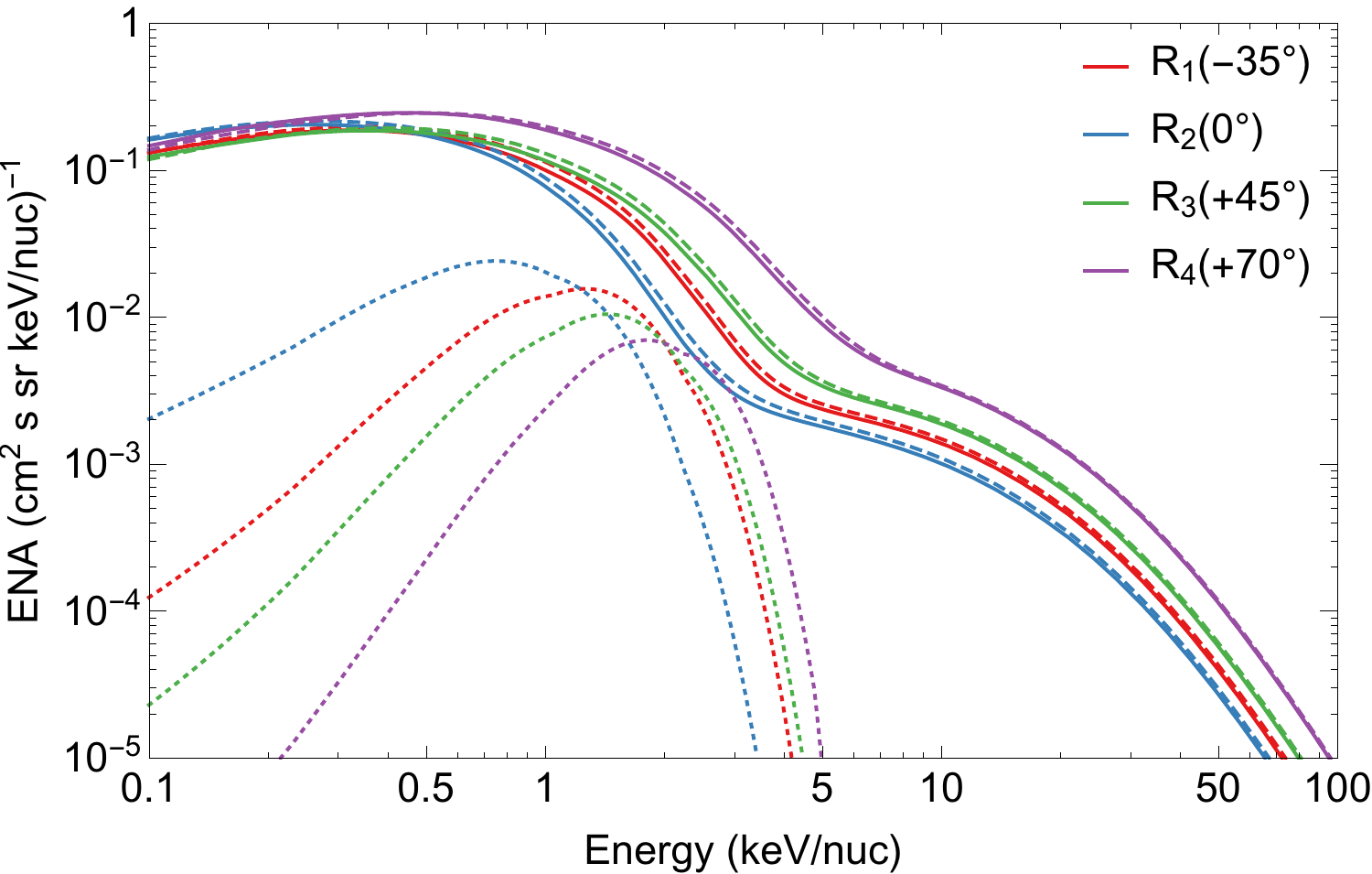}

  \plotone{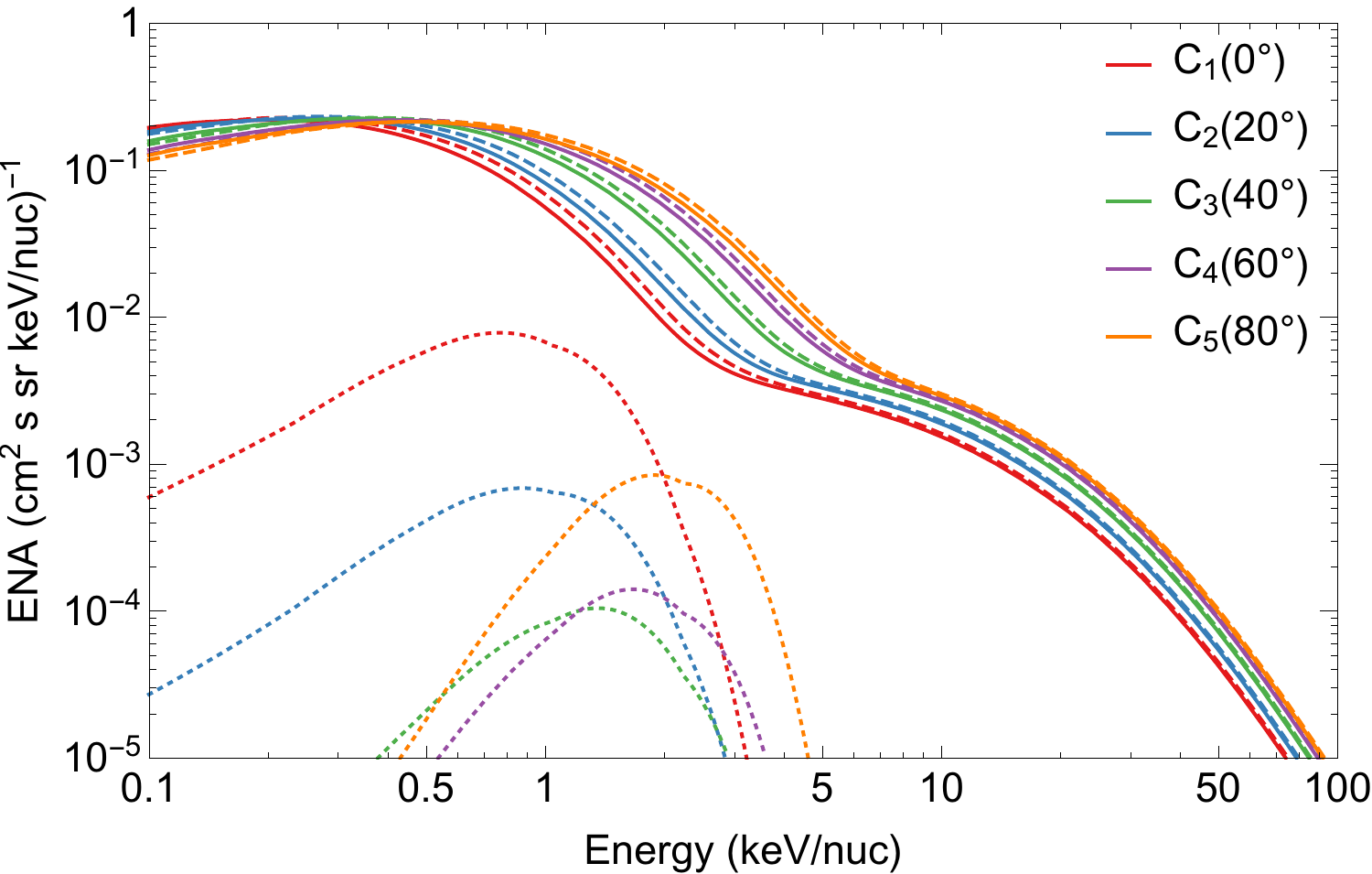}

  \caption{Energy spectra of He ENA expected at 1 au. The solid and dashed lines present the signal expected in the model with the potentials $\Phi_1$ and $\Phi_2$, respectively. The ribbon component is presented with the dotted lines; results from both flow potentials are indistinguishable in the scale of this drawing. \label{fig:spectraheena}}
\end{figure}

Figure~\ref{fig:spectraheena} presents the energy intensity spectra of He ENAs in some selected directions. The solid and dashed lines present the total intensity from both components in these directions for the flow resulting from the potential $\Phi_1$ and $\Phi_2$, respectively. Separately presented is the contribution of the ENAs created in the secondary ENA mechanism (dotted line). This secondary ENA contribution was calculated separately for these two potentials. The only difference between them is due to a different distance to the heliopause. However, the resulting difference in the ENA signal is negligible and is not noticeable in the scale of these spectra. 

The obtained spectra have characteristic shapes resulting from the assumed shape of the PUIs spectrum at the termination shock. The roll-over of the spectrum, observed between 0.2 and 0.6 keV/nuc, is due to the maximum of the PUIs distribution shifted to the lower energies because of the plasma flow in the inner heliosheath. This effect may not be visible in the actual observations because some potential ENA sources important for energies $\lesssim0.5$~keV/nuc are neglected in this analysis. They include the newly born PUIs in the inner heliosheath, or heated plasma in the outer heliosheath. For higher energies the spectrum is monotonic, with some flattening at $\sim$5--10 keV/nuc due to the contribution of the reflected PUIs (compare Figure~\ref{fig:tsspectrum}). This result suggests that observation of He ENA allows for indirect insight into the acceleration of helium ions at the termination shock.

The top panel of Figure~\ref{fig:spectraheena} presents the spectra in the upwind direction (Nose), the downwind direction (Tail), in the directions of \emph{Voyager 1} (V1) and \emph{Voyager 2} (V2), and in the directions shifted northward (NT) and southward (ST) from the downwind direction by an angle of 10$\degr$. Analysis of this figure shows that the tail-to-nose ratio reaches about 100. Additionally, the signals in the nose and in the \emph{Voyager} directions are close to each other. The signal for the potential $\Phi_2$ compared to that for the potential $\Phi_1$ is systematically larger for all directions, but the difference is small (except for the downwind part). This is explained by the adiabatic heating of the plasma due to the slowdown caused by charge exchange. 

Larger differences are in the tail directions. For the flow $\Phi_1$ without slowdown due to charge exchange, the signals in the ST and NT directions are larger than in the tail for energies $\sim$1--4 keV/nuc. This effect is not present for the flow $\Phi_2$ with the slowdown. The differences between these potentials in the downwind side are due to bending of the flow lines toward the downwind direction in the potential $\Phi_2$. Consequently, the lines of sight in this part of the sky cross the heliopause at smaller distances from the Sun, i.e., the integration paths are shorter. Simultaneously, they cross the flow lines originated at higher heliographic latitudes at smaller distances, where the density of helium ions is not significantly reduced due to charge exchange. For the potential $\Phi_1$, these lines of sight cross the flow lines originating at the higher heliographic latitudes farther in the tail, where energetic helium ions are almost extinct.

The middle panel shows spectra in selected directions along the \emph{IBEX} ribbon. It illustrates that the component of the secondary ENAs from the outer heliosheath is comparable to the inner heliosheath contribution, but not dominant, as it is observed in these directions for hydrogen \citep{schwadron_2014a}. The directions along the ribbon are presented in the maps and were chosen to have different heliographic latitudes: $-35\degr$, $0\degr$, $+45\degr$, and $+70\degr$. Both components of the signal show that with increasing latitudes, the spectra shift to higher energies. For the inner heliosheath emission, this comes out because of the assumed energies of the transmitted PUIs at the termination shock (see Section~\ref{sec:spectra}) as a function of latitude. For the secondary ENA components, the flux of the primary ENAs originating in the supersonic solar wind shows a similar dependency.

\begin{figure*}
  \epsscale{1.15}
  \plottwo{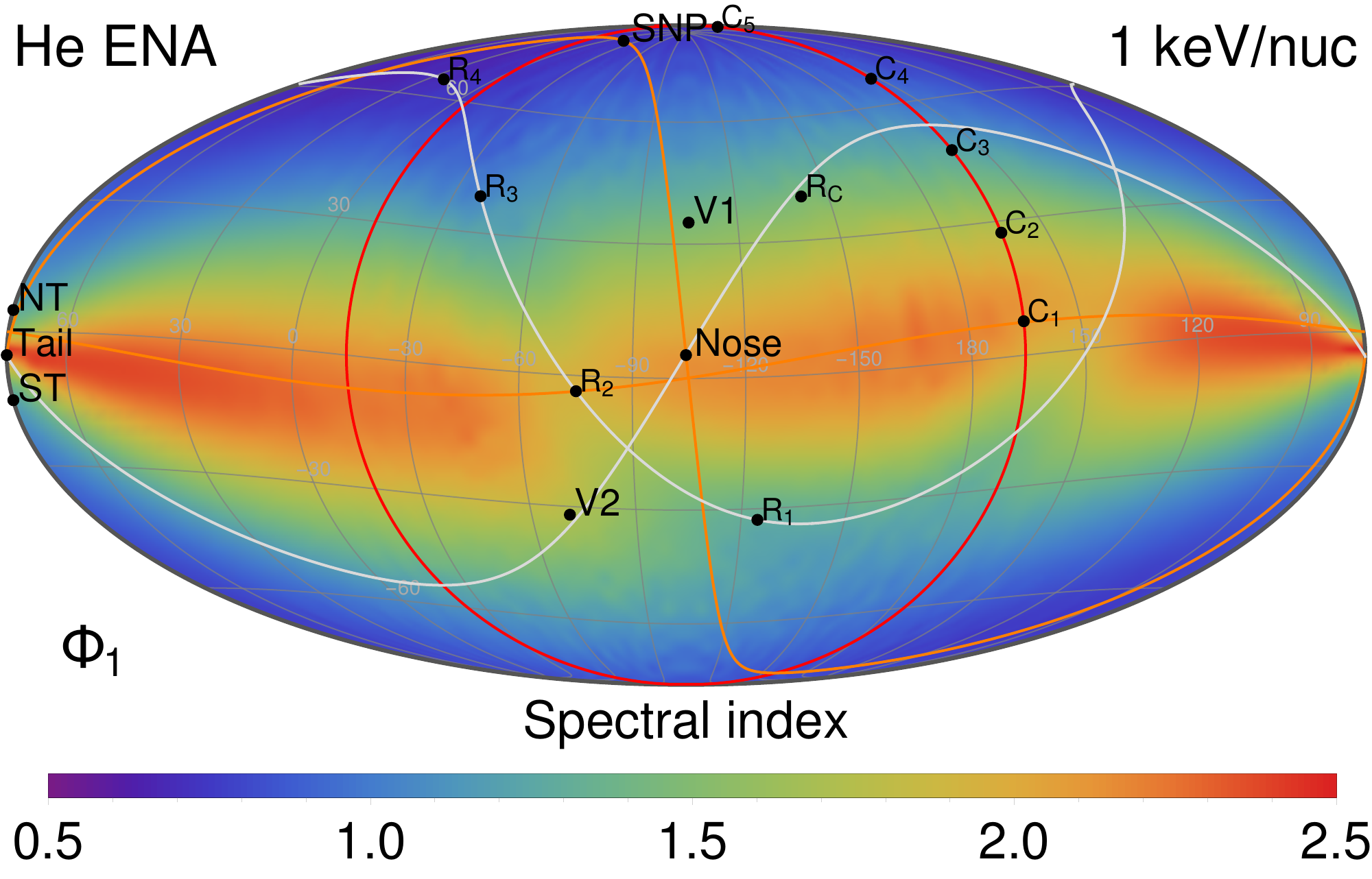}{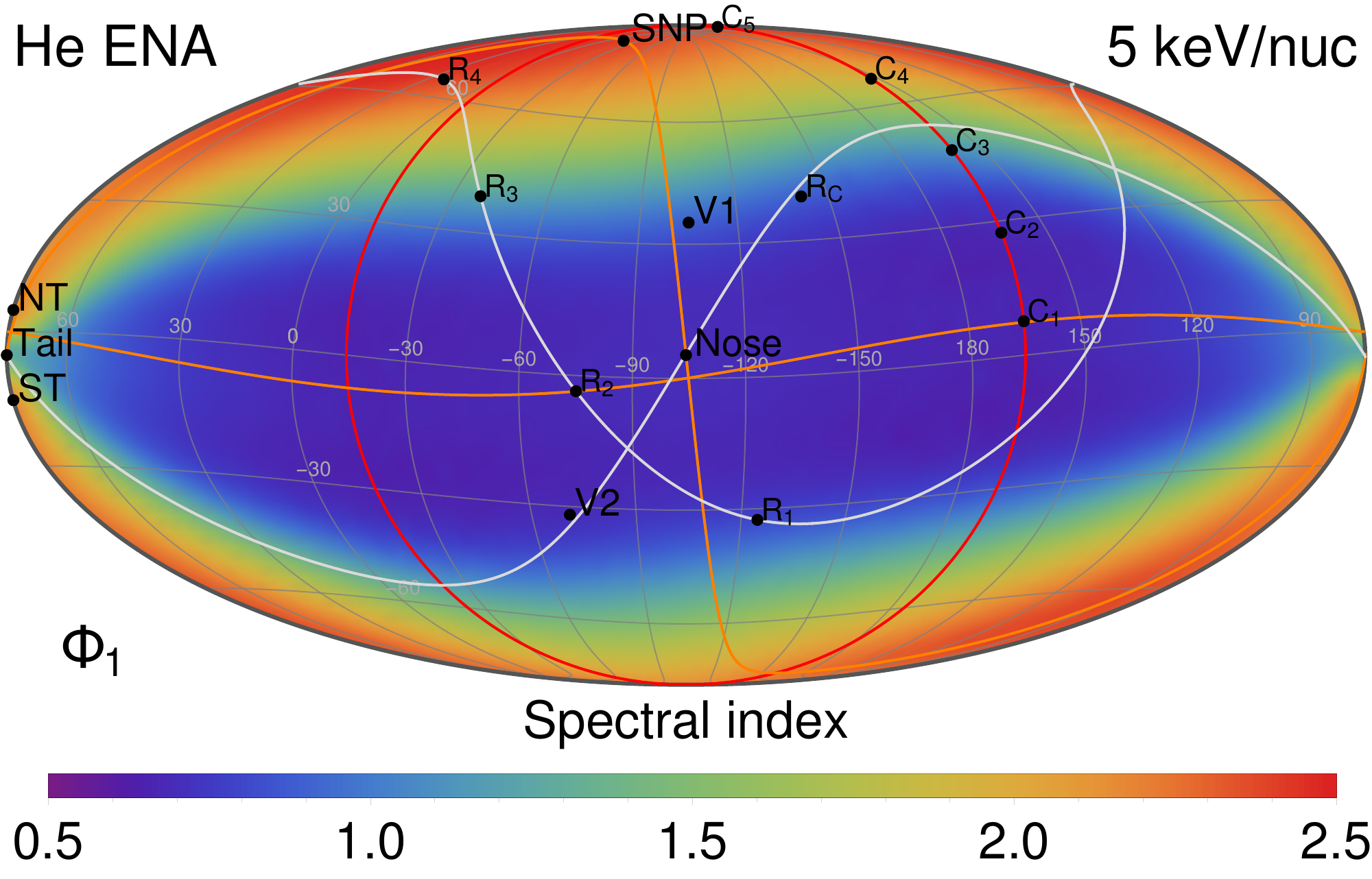}
  \caption{Spectral indices of the expected He ENA signal at 1 keV/nuc (left panel) and 5 keV/nuc (right panel) with the flow from the potential $\Phi_1$. The spectrum for 1 keV/nuc is harder at high latitudes and softer at lower. For the energy 5 keV/nuc the situation is opposite.\label{fig:spectralindex}}
\end{figure*}

The lowest panel, which presents the spectra in the crosswind directions, also confirms this observation. The directions are chosen so that they all are directed exactly perpendicular to the upwind direction but have different heliographic latitudes. Figure~\ref{fig:spectralindex} presents maps of the spectral indices given by the formula
\begin{equation}
 \gamma=-\frac{\partial \log j_\mathrm{HeENA}}{\partial \log E}
\end{equation}
at two energies of 1~keV/nuc and 5~keV/nuc. At lower latitudes, the spectral index is larger at 1 keV/nuc and smaller at 5 keV/nuc because of the flattening of the spectrum. For larger latitudes, the spectrum is harder at 1 keV/nuc due to the shifted maximum of the signal, but at the 5 keV/nuc the spectrum is not yet flattened, and thus the spectral index is larger.

\section{Discussion}\label{sec:discussion}
Motivation for development of an ENA detector capable of observing helium should arise from specific scientific questions that can be answered based on this observation. In Section~\ref{sec:imap}, technical requirements for the construction of such a detector are discussed, including the necessary geometrical factor to observe He ENAs from the heliosphere. Further on, reasons and implications of the hourglass-shaped signal from the downwind direction are discussed (Section~\ref{sec:hourglass}), as well as the long mean free paths of helium ions in the heliosphere resulting in the large tail-to-nose signal ratio (Section~\ref{sec:tail}). 

Besides He ENAs from the heliosphere, some ENAs from extraheliospheric sources can be expected. \citet{swaczyna_2014} showed that with the extraheliospheric hypothesis of the \emph{IBEX} ribbon \citep{grzedzielski_2010}, the intensity of He ENAs from the neighboring boundary layer of the Local Interstellar Cloud and the Local Bubble is much higher than the heliospheric signal. Regardless of the plausibility of this hypothesis, this result suggests that if due to some reason the plasma in the proximity of the heliosphere contains suprathermal ions, they can be detected using the resulting He ENAs. This is possible because the mean free path of He ENAs in the LISM is much longer than that of hydrogen. Detailed discussion of such sources is out of the scope of this paper. 

\subsection{Requirements for ENA detectors}\label{sec:imap}
He ENAs from the heliosphere are a good target for a new generation of ENA detectors. The \emph{IMAP} spacecraft will be equipped with two ENA detectors \citep{nrc_2013}. They will be able to observe He ENA if their geometric factors are high enough and they are equipped with a mass spectrometer.

The first requirement can be translated as the requirement for the number of counts expected to be observed per energy channel and pixel during some time of operation. \citet{funsten_2009} defined the energy geometric factor $G^\mathrm{E}$, which connects the ENA intensities $j_\mathrm{ENA}$ with the number of observed counts $n$ by the equation
\begin{equation}
 j_\mathrm{ENA}=\frac{n}{\Delta t E G^\mathrm{E}}, 
\end{equation}
where $\Delta t$ is the time of observation per pixel and energy channel and $E$ is the central energy of the channel. This equation can be rewritten as:
\begin{equation}
 G^\mathrm{E}=\frac{n}{j_\mathrm{ENA}\Delta t E}. \label{eq:egeomfact}
\end{equation}
Based on this equation it is possible to calculate the minimum required energy geometric factor that allows for observation of He ENAs from the heliosphere. In this section, the time of observation is adopted as:
\begin{equation}
 \Delta t=\frac{t_\mathrm{tot}}{n_\mathrm{pix}n_\mathrm{ch}},
\end{equation}
where $t_\mathrm{tot}=0.5$~year is an effective time of the observation cycle, $n_\mathrm{pix}=1800$ is the number of pixels, and $n_\mathrm{ch}=6$ is the number of energy channels. The adopted values correspond to the observational strategy of \emph{IBEX}-Hi instrument. The effective time of 0.5 year corresponds to 1 year of observations needed to obtain one full ram map. Effectively, this time is shorter for \emph{IBEX} because of the limitation of good time intervals, but most of the reasons for these limitations should be absent for \emph{IMAP}, which will be located in the L1 Sun-Earth Lagrangian point. The number of pixels corresponds to the $6\degr \times 6\degr$ pixels used for analysis of \emph{IBEX} data. The number of energy channels is also assumed as for \emph{IBEX}-Hi. With these numbers, the time of observation per pixel and energy channel is 1460 seconds during 1 year. 

\begin{figure}
  \epsscale{.6}
  \plotone{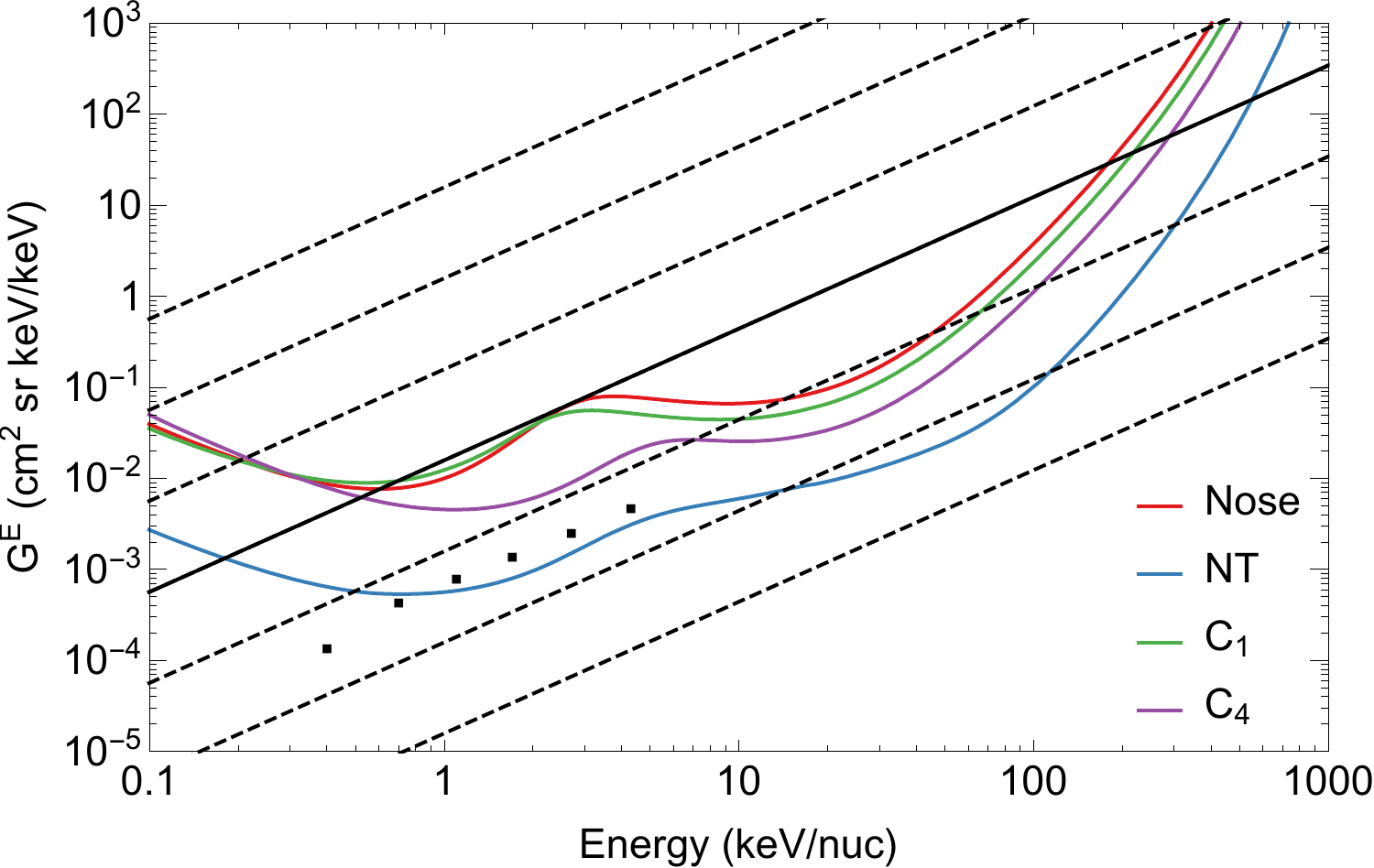}
  \caption{Minimum required energy geometric factor for observation of one count per pixel and energy channel in the selected directions in the sky (color solid lines). The expected geometric factor for helium is marked with the black solid line, assumed to be equal 30 times the energy geometric factor for hydrogen on \emph{IBEX}-Hi (black squares). The oblique dashed lines present multiplications of the expected geometry factor with powers of 10, i.e., they connect points with the same number of expected counts (e.g., the first line below black marks 10 counts).\label{fig:gf}}
\end{figure}

Figure~\ref{fig:gf} presents the results of Equation~\eqref{eq:egeomfact} applied to the spectra of He ENAs in four directions for the flow described by potential $\Phi_1$: the upwind direction, two in the crosswind directions, and in the north tail direction. It is compared with the energy geometric factor for hydrogen for the \emph{IBEX}-Hi energy channels. \emph{IBEX}-Hi detects the incoming ENAs owing to the ionization of the incoming atoms on thin carbon foils. This process is less efficient for helium than for hydrogen. \citet{allegrini_2014} showed that the ionization of helium atoms on the carbon foil for the same energy per nucleon is about 3 times less probable than for hydrogen. However, the detectors on the \emph{IMAP} spacecraft should have about 80 times larger efficiency \citep{nrc_2013}. The figure presents the reference level of the expected energy geometric factor for helium assumed to be $\sim$30 times larger than the current level for triple coincidences in \emph{IBEX}-Hi. 

The figure shows that with such a reference level of the energy geometric factor, the observations of He ENAs are feasible. In the downwind direction, where the signal is the strongest, the number of counts per pixel and energy channel can reach 10 in the energy range 1--200 keV/nuc. The number of counts is larger than 1 between 5 and 100 keV/nuc in all of the directions. The largest number of counts is expected for a few tens of keV/nuc. 

The expected count numbers observed by \emph{IBEX}-Hi per pixel and energy channel based on the 5 year data \citep{mccomas_2014b} range from $\sim$20 to $\sim$200 per year. With the increased efficiency on \emph{IMAP} this means that in some part of the sky it would reach about 2000--20000 counts per pixel and energy channel in one year. Simultaneously, the expected number of helium counts is 1--200. Consequently, the mass spectrometer on the ENA detector must be specialized, so that a false helium count caused by an impact of a hydrogen atom is less probable than about 1 in $10^5$, i.e., it is at least one order of magnitude better than the typical ratio of counts. For the time-of-flight spectrometer this is feasible due to the twofold difference in the atom speeds for the same total energy (not energy per nucleon). This high ratio of atoms shows why the idea of a mass analysis technique using coincidence ratio in the \emph{IBEX}-Hi instrument, proposed by \citet{allegrini_2008}, is not sufficient for helium. 

This section provides an estimate on the possibility of observation of He ENAs based on the properties of the desired capabilities of the ENA detectors on \emph{IMAP}. The count numbers expected in each $6\degr\times6\degr$ pixel can be enhanced by integration over a longer time or on larger pixels. However, the increase in the resolution requires some reduction of the detector field-of-view. This would lead to a decrease in the energy geometric factor and thus reduce the helium signal. Consequently, it is important to find an optimal balance between the resolution of hydrogen observations and the ability to observe helium.

\subsection{An hourglass-shaped signal in the downwind hemisphere} \label{sec:hourglass}
The maps of the intensities of He ENAs show an hourglass shape of the signal in the downwind hemisphere. This structure is most pronounced for energies 1--3 keV/nuc. The hourglass effect is caused by the latitudinal structuring of the supersonic solar wind. A line of sight in any direction other than within the solar equatorial band crosses multiple flow lines that originate at different heliographic latitudes at the termination shock. The flow lines crossed by this line of sight feature the evolved PUI spectra characteristic for different heliographic latitudes and the resulting ENA spectrum has contributions from all of them. Consequently, the intensity of He ENAs from higher latitudes is larger than the signal from lines of sight at similar distances from the upwind direction but crossing lower latitudes (see the lowest panel of Figure~\ref{fig:spectraheena}). However, due to the long mean free path against neutralization of helium ions in the inner heliosheath (see Section~\ref{sec:tail}) and thus the high tail-to-nose ratio, the signal in the downwind hemisphere shows a complex structure.

\begin{figure}
  \epsscale{.6}
  \plotone{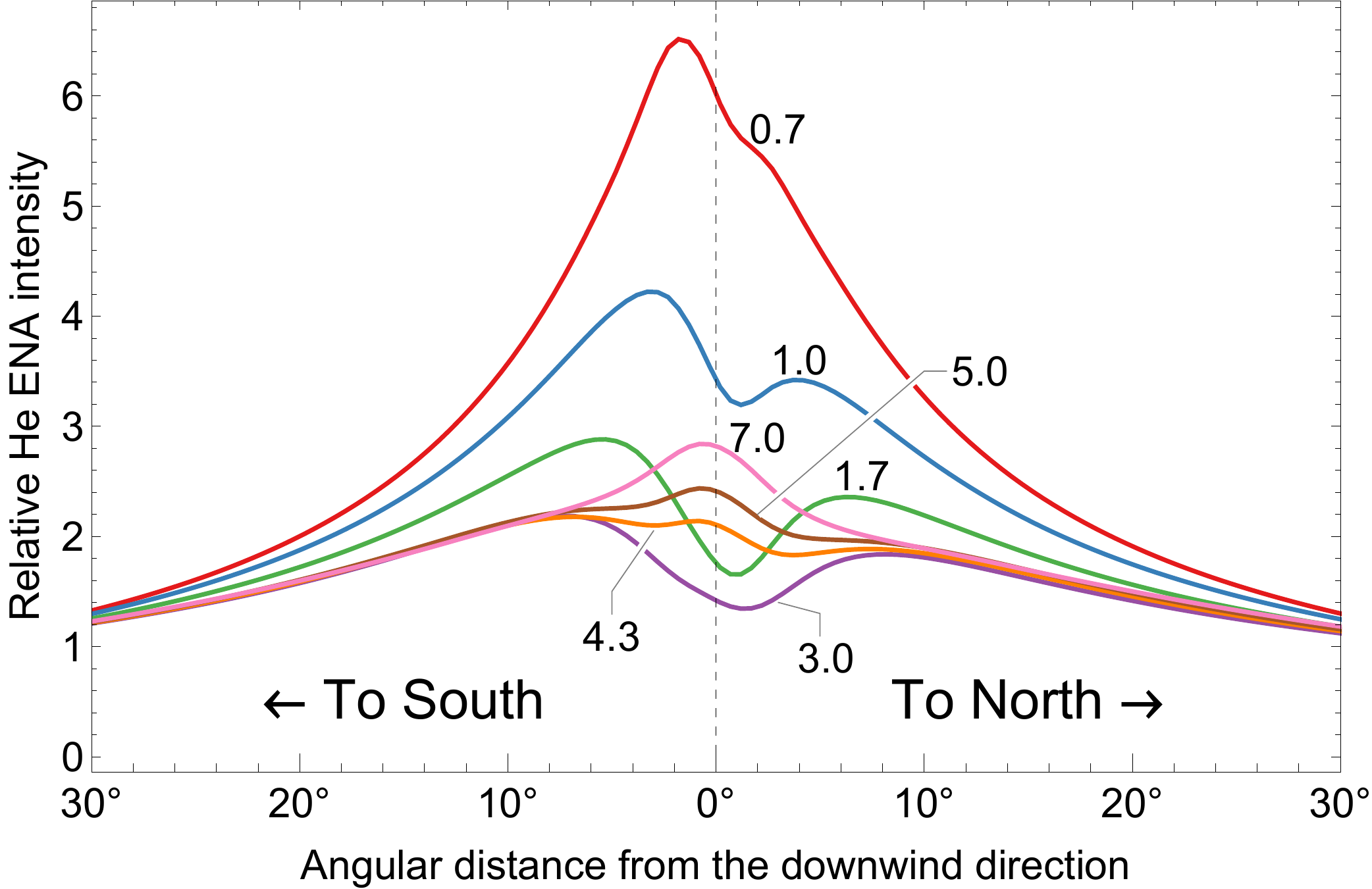}
  \caption{Normalized intensities for potential $\Phi_1$ along the great circle (meridian) crossing the poles and the downwind direction. Note that for the energies 1--4 keV/nuc the signal has two peaks away from the downwind direction. \label{fig:hourglass1}}
\end{figure}

\begin{figure}
  \epsscale{.6}
  \plotone{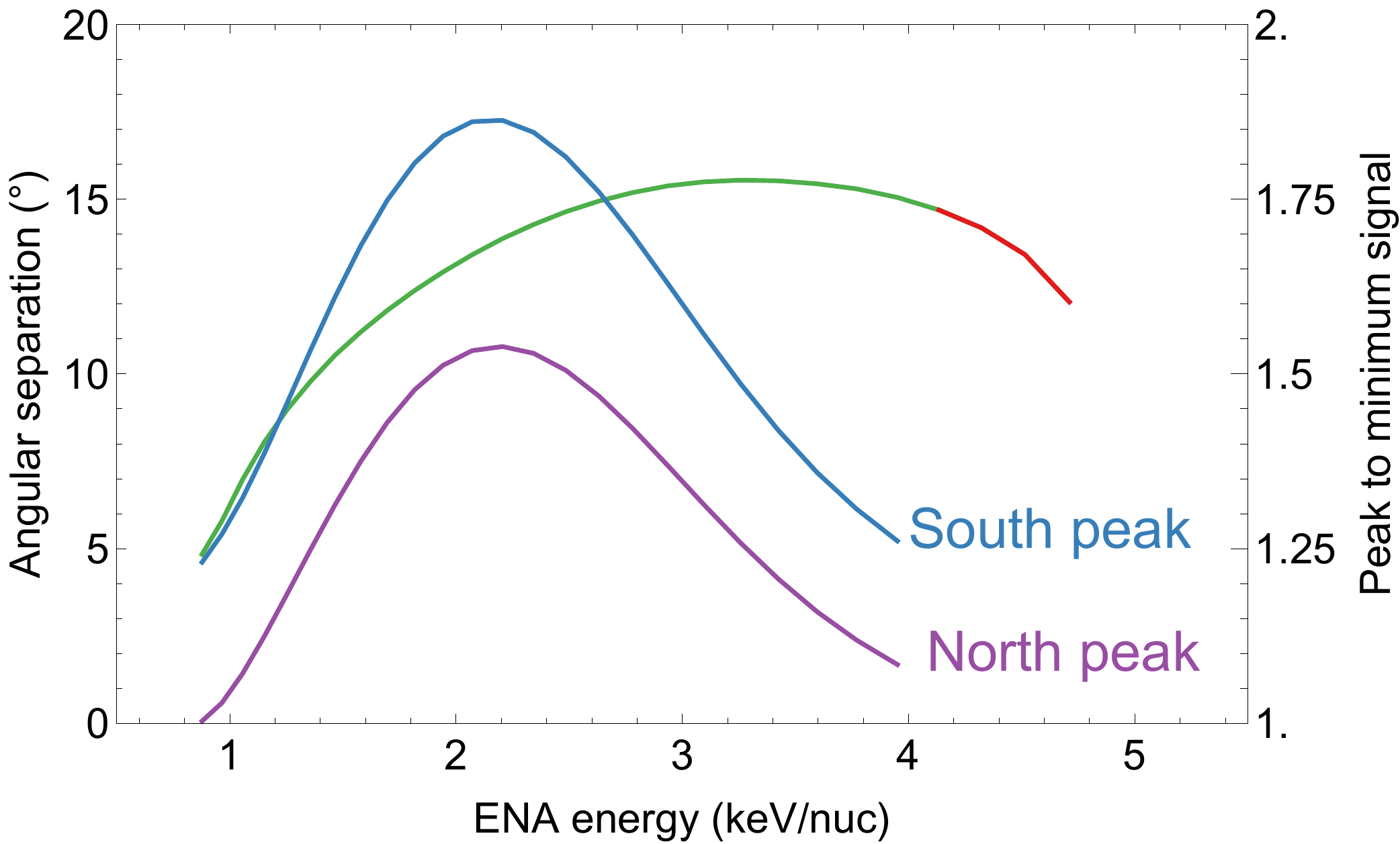}
  \caption{Angular separation of the peak positions (green line, left scale) and the ratio of the signal at the peak to the minimum between the peaks (right scale). The red portion of the separation line correspond to the range with three peaks (cf. text).\label{fig:hourglass2}}
\end{figure}

For energies $\sim$1--4 keV/nuc the maps obtained for the flow due to potential $\Phi_1$ have two local maxima. These maxima are located at the heliographic meridian that crosses the downwind direction. Figure~\ref{fig:hourglass1} presents the relative intensities of He ENAs along this meridian for some of the energies between 0.7 and 7 keV/nuc. From this figure it is clear that for certain energies more than one peak is present. Figure~\ref{fig:hourglass2} shows the angular separation of these peaks as a function of energy and the contrast of the peaks, defined as the ratio of the peaks to the minimum value between them. The largest separation between the peaks is for $\sim$3.2 keV/nuc.

The analysis of intensities along the meridian shows that for energies $\gtrsim$4~keV/nuc, an additional third peak is visible in the downwind direction. The peak separation for the edges of the energy range where at least two peaks are visible does not reduce to zero because the local maxima change into inflection points rather than merging into one peak. The peak located southward from the downwind direction is higher because the downwind direction is located closer to the Sun's South Pole. The heights of the peaks relative to the minimum signal for the given energy and the magnitude of the angular separation of the peaks suggest that these features can be observed with the ENA detector on \emph{IMAP} at least for energies $\sim$2--3~keV/nuc. However, the collimator transmission function may smooth out these features for lower and higher energies because of the small contrast of the peaks.

The two maxima featured in the downwind hemisphere are not visible for the flow given by potential $\Phi_2$. With this flow, the cross section of the tail is decreasing with the solar distance and the flow lines are bent toward the downwind direction. In this case, for all of the analyzed energies, the maps have a single maximum exactly at the downwind direction.

\citet{zirnstein_2016b} quantitatively described a similar structure in the \emph{IBEX}-Hi observations in the two highest energy steps. They also calculated the hydrogen ENA emission from similar models to the models used in this analysis. However, in their analysis, it was assumed that the PUI distribution functions at the termination shock do not depend on the heliographic latitude, and thus the ENA intensities have an axial symmetry. Recently, \citet{zirnstein_2017} calculated the hydrogen ENA emission in the time-dependent magnetohydrodynamic model, which includes the helio-latitudinal structure of the solar wind. They showed that partitioning of the plasma energy into three populations of protons: solar wind, transmitted PUIs and reflected PUIs, allows for reproduction of the split tail feature in the fluxes of hydrogen ENAs. Other models of hydrogen ENA emission that include the latitudinal structure of the solar wind \citep{zank_2010, zirnstein_2014, czechowski_2015a} also reproduce this effect qualitatively.
	
In this analysis, the effect of helio-latitudinal structure of the solar wind is included partially. This structure is reflected in the energies of PUIs in the inner heliosheath, but it does not affect the plasma flow in the inner heliosheath. Generally, both factors can impact the resulting ENA production. The flow with the structure included would affect the ENA intensities due to different densities, velocities in the Compton--Getting effect, and evolution of helium ions. However, \citet[][Section 4]{zirnstein_2017} presented that the separation of the split tail depends strongly on the PUIs energy rather than on the assumed flow velocities.

\subsection{Mapping the tail structure}\label{sec:tail}
\begin{figure}
  \epsscale{.6}
  \plotone{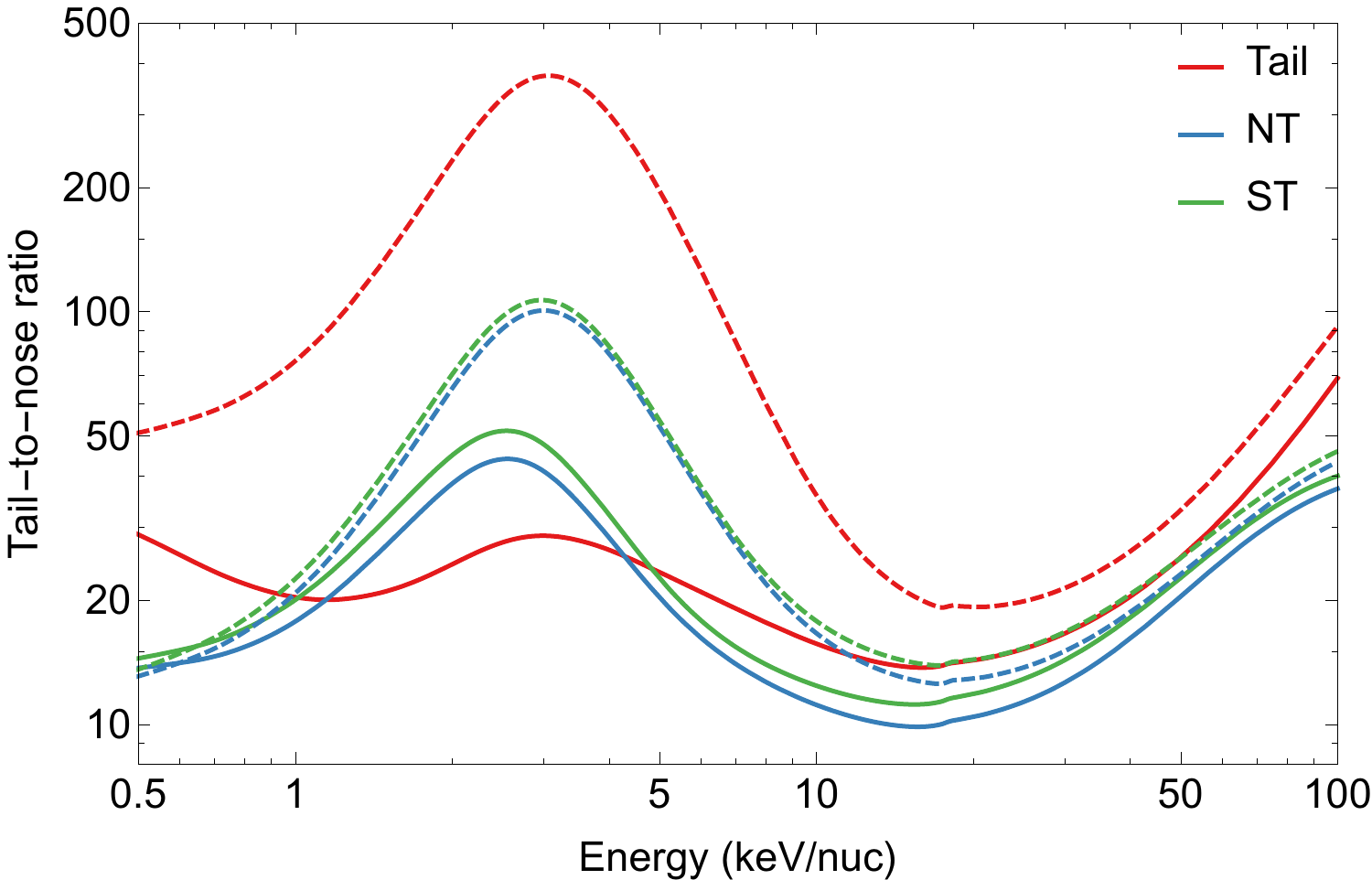}
  \caption{Ratios of the intensity spectra of He ENAs in the directions located in the heliospheric tail to those in the upwind direction. The results given by the flow from the potential $\Phi_1$ ($\Phi_2$) are plotted with solid (dashed) lines.\label{fig:tano_ratio}}
\end{figure}

The intensities of He ENAs show a large nose to tail ratio. Figure~\ref{fig:tano_ratio} shows the ratio of the intensity spectra in the Tail, NT, and ST directions to those for the upwind direction for various energies. This ratio is much larger than the range of hydrogen ENA intensity ratio observed by \emph{IBEX}. 

These high ratios can be explained by the deeper penetration in the inner heliosheath of helium ions than protons. The mean free paths against neutralization of helium ions in the inner heliosheath against various processes are discussed in Appendix~\ref{appendix:crosssec}. The mean free path at 1 keV/nuc can extend to about 1000~au along the flow line. For energies of about 20 keV/nuc this means that free paths decrease to about 200~au and the resulting tail-to-nose ratio is smaller. 

Some of the assumptions made in this analysis may have a certain influence on the obtained intensities of He ENAs. One of them is that the forms of the potentials used here assume that the termination shock is spherical. Global magnetohydrodynamical models of the heliosphere show that the distance to the termination shock in the downwind direction can be up to $\sim$2 times larger than in the upwind direction \citep[e.g.,][]{pogorelov_2009}. This analysis considers only a spherical termination shock; however, inclusion of a non-spherical termination shock in the flows derived from a scalar potential is possible \citep{zirnstein_2016b}. The asymmetry of the termination shock may influence presented results. First, the accumulated fluxes of helium PUIs created in the supersonic solar wind in the downwind side can be larger due to longer accumulation path. Moreover, the intensities of He ENAs may change due to modified lower integration limit set at the termination shock.

The other assumption is the constant density of the background interstellar neutrals in the inner heliosheath. The global multi-component models of the heliosphere show that the densities are up to 2 times smaller in the downwind direction than in the upwind or crosswind directions \citep{muller_2008, heerikhuisen_2015}. Consequently, the production rate of He ENAs in the downwind side are likely overestimated in this analysis, and in the reality the ENA fluxes may be smaller.

The effects of these two assumptions tend to cancel each other out. The spherical termination shock assumed in this analysis should tend to decrease the tail-to-nose ratio, while the assumption of the constant density of neutral hydrogen and helium should increase this ratio. Consequently, the joint effect on the estimation of the intensities of He ENAs should be generally smaller than their partial contributions.

For the lower energies, the decrease in the tail-to-nose ratios is caused by two effects. One of them is the Compton--Getting effect, which modifies the spectra of emitted ENAs more effectively for lower energies. In the upwind region, the slowdown of the plasma flow reduces the difference between the ENA velocities in the Sun's frame and those of parent ions in the plasma frame, compared to the downwind region, where the outward flow is approximately constant (see Figure~\ref{fig:plasmaflow}). The other effect is that ENAs at lower energies are longer exposed to the ionization processes.

The shape of the heliosphere is a subject of discussion in the heliospheric community. The considered shapes are a comet-like extended tail, which is used in this analysis \citep[e.g.,][]{parker_1961}, a flattened bubble without a tail \citep[e.g.,][]{krimigis_2009}, and a croissant-shaped heliosphere with two lobes, extending in two directions \citep[e.g.,][]{opher_2015}. 

\citet{schwadron_2014a} noticed that energetic protons in the inner heliosheath are neutralized at distances of $\sim$100~au. Consequently, almost all of PUIs created at the termination shock are neutralized before they progress deeper in the heliospheric tail. Eventually, the distinction between different models of the heliospheric tail may not be possible only by observing the hydrogen ENAs if the expected fluxes in each of these models are comparable. However, observation of He ENAs may help resolve the dilemma about the shape of the heliospheric tail. In this analysis, a solution for the comet-like tail heliosphere is presented. For the bubble-shaped and the croissant-shaped heliospheres, the heliopause in the downwind direction should be relatively close, and consequently the signal from this direction should not be larger by the factor $\gtrsim$10 obtained in the comet-shaped model. For the croissant-shaped heliosphere, the highest signal is expected to be aligned with the two distant tails for most energies, whereas for the bubble-shaped heliosphere the inner heliosheath signal in different directions should be comparable, except perhaps in the direction aligned with the interstellar magnetic field.

\section{Conclusions}\label{sec:conclusions}
This paper presents estimates of the expected intensities of He ENAs from the inner and outer heliosheath. The feasibility of observations of these atoms with future ENA detectors is presented. The results are discussed in the context of the some poorly understood properties of the heliosphere that can be resolved using observations of He ENAs.

Two simple models of the plasma flow in the inner heliosheath were selected for calculation of the evolution of helium ions in the inner heliosheath and production of He ENAs. In these models, the flows are given as a gradient of scalar potentials. The model parameters were adopted based on the observations of \emph{Voyagers}. The models include magnetic fields neither in the inner heliosheath nor in the interstellar medium. They also assume that the plasma is incompressible and irrotational. The resulting flows have rotational symmetry around the direction of the Sun velocity with respect to the interstellar medium.

The distributions of helium ions in the inner heliosheath are calculated by evolving their spectra from the termination shock along the flow lines. The termination shock spectra account for the latitudinal structure of the solar wind. The distributions of the helium ions are subsequently used to integrate the production of He ENAs along various lines of sights, intersecting the inner heliosheath (Section~\ref{sec:ihs}). 

The outer heliosheath contribution to He ENAs is calculated from an analytic model of the secondary ENA mechanism. This contribution is smaller than the inner heliosheath signal in all directions in the sky (Section~\ref{sec:ohs}). Consequently, the \emph{IBEX} ribbon should not be as pronounced in the helium observations as it is for hydrogen if it is produced in the secondary ENA mechanism. 

The absolute intensities of He ENAs are small compared with hydrogen ENAs. The typical intensities for 1~keV/nuc range from 0.05 to 2~$\mathrm{ENA\,(cm^2\,s\,sr\,keV/nuc)^{-1}}$, and for 10~keV/nuc range from 0.001 to 0.015~$\mathrm{ENA\,(cm^2\,s\,sr\,keV/nuc)^{-1}}$ (Section~\ref{sec:results}). Nonetheless, observations of He ENAs should be possible using the next generation of ENA detectors with higher efficiencies, as those expected for the ENA detectors on the planned \emph{IMAP} mission (Section~\ref{sec:imap}). Successful determination of He ENAs requires, however, providing these detectors with a mass spectrometer capability to allow for identification of chemical elements. 

The estimated signal in the downwind side has an hourglass shape approximately aligned with the heliographic meridian (Section~\ref{sec:hourglass}). This effect results from a combination of the long integration path in the tail directions and the inclusion of the solar wind latitudinal structure in the PUI distribution functions. 

The intensities of He ENA show a high nose-to-tail ratio, ranging from ten to more than a hundred, depending on the ENA energy (Section~\ref{sec:tail}). This is a result of the long mean free path against neutralization of helium ions in the inner heliosheath, which for energies of $\sim$1~keV/nuc exceed 1000~au along the flow lines. As a consequence, the helium ions are more homogeneously distributed in the inner heliosheath than protons. The signal is effectively integrated along a longer path along the lines of sight that have long distances to the heliopause. Since protons emerging from the termination shock are neutralized on shorter paths and thus do not penetrate the distant part of the heliospheric tail, such an effect is not visible.

The high signal of He ENAs from the heliospheric tail can be used as a promising tool to distinguish between competing models of the heliospheric tail. This analysis assumed that the heliosphere has a single tail in the direction opposite to the Sun's motion. However, for models with two lobes \citep{opher_2015}, it is expected to have maxima in the direction of the lobes, so the enigma of the heliospheric shape should be resolved once the first global map of the sky observed in He ENAs is obtained. 

\acknowledgments

The authors acknowledge support by the grant 2015/19/B/ST9/01328 from the National Science Centre, Poland.

\appendix

\section{Charge-change reactions for helium}\label{appendix:crosssec}
The list of the essential binary interactions (processes) for helium is longer than that needed for analysis of hydrogen ENAs. The relevance of a process can be expressed by its reaction rate. The reaction rate is a product of the cross section $\sigma$, the relative velocity $v$, and the density of the second reagent $n$: $\alpha=\sigma v n$. Typically, reactions of ions and atoms of the same element have larger cross sections. Additionally, hydrogen and helium are the dominating elements in space. Consequently, the important reagents for the reactions with helium ions and atoms are hydrogen atoms (H), protons (H$^+$), electrons (e$^-$), helium atoms (He$^0$), singly ionized helium ions (He$^+$) and doubly ionized helium ions (He$^{2+}$). 

In general, the reaction rate should be calculated as an integral of a product of the cross section, the relative speed, and the distribution function over the entire velocity space of the second substrate. It is assumed for the second substrate that they follow Maxwell--Boltzmann distribution, also typically their thermal speeds are small compared to the velocity of the first reagent. Under these assumptions, the spread of the relative velocity is small, and the simplified formula is justified \citep{heerikhuisen_2015}. Only for electrons the thermal speed is comparable with the speeds of analyzed helium ions and atoms, and thus for reactions with electrons the integral was calculated directly.

This appendix considers the reactions that change the charge state of helium and identifies those that are important for the analysis presented in this paper. The reactions that do not change the charge state are neglected. 

The speed of helium ions or atoms considered in the analysis is typically much higher than the flow speed or thermal speed of the second reagent. Consequently, the rates are presented as a function of projectile energy in this appendix. However, in the case of reactions with electrons, the velocities of electrons are much larger than these of atoms and ions due to the high mass ratio, and in this special case the thermal speed of electron is added to the ``projectile'' velocity of helium. 

The contribution of different reactions are compared with each other using the effective mean free paths of helium ions against each of the reaction given by
\begin{equation}
 \lambda=\frac{u}{\alpha},
\end{equation}
where $\alpha$ is the reaction rate and $u$ is the flow speed. For ENAs the formula is
\begin{equation}
 \lambda=\frac{v_\mathrm{ENA}}{n\sigma v_\mathrm{rel}},
\end{equation}
where $v_\mathrm{ENA}$ is the ENA velocity, and $v_\mathrm{rel}$ is the relative speed. For energetic atoms, typically $v_\mathrm{ENA}=v_\mathrm{rel}$, because the ENA velocity is much larger than the flow speed and the thermal speed. However, for electrons, their speeds need to be accounted for due to the high mass ratio. The combined mean free path for several reactions can be calculated as the harmonic sum of the mean free paths characteristic for these individual reactions. 

Table~\ref{tab:appendixreaction} presents all reactions that are included in this analysis. In the subsequent sections, reactions that change the charge-states of He$^{2+}$ (Section~\ref{sec:he2}), He$^{+}$ (Section~\ref{sec:he1}), and He$^{0}$ (Section~\ref{sec:he0}) are presented. In all parts of the analysis, the reaction rates were calculated according to the actual densities of the reagents as measured for the specific points. However, in this appendix we compare mean free paths for the reactions using the following constant values of densities: H$^0$ -- 0.1~cm$^{-3}$, H$^+$ -- 0.002~cm$^{-3}$, He$^0$ -- 0.015~cm$^{-3}$, He$^+$ -- 0.00005~cm$^{-3}$, He$^{2+}$ -- 0.0001~cm$^{-3}$, and e$^-$ -- 0.00225~cm$^{-3}$. These values correspond to the typical abundances of the particles in the inner heliosheath. The value of the flow speed is adopted as 150~$\mathrm{km\,s}^{-1}$. 

\begin{deluxetable*}{rllcC}
  \tablecaption{Reactions modifying the charge state of helium\label{tab:appendixreaction}}
  \tabletypesize{\footnotesize}
  \tablehead{
	\colhead{} & 
	\colhead{Symbol\tablenotemark{a}} & 
	\colhead{Formula} &
	\colhead{Ref.} &
	\colhead{Importance\tablenotemark{b}}
  }
  \startdata
  \multicolumn{5}{c}{$\mathrm{He^{2+}\to He^+\;or\;He^0}$} \\
  \hline
  1 & cx-He$^{0}$ 	& $\mathrm{He^{2+}+He^0\to He^+ + He^+}$ 			& (1) & ++\\
  2 & cx-He$^{+}$	& $\mathrm{He^{2+}+He^+\to He^+ + He^{2+}}$ 		& (1) & \\
  3 & cx-H$^{0}$ 	& $\mathrm{He^{2+}+H^0\to He^+ + H^+}$ 				& (2) & ++\\
  4 & rc-e$^{-}$ 	& $\mathrm{He^{2+}+e^{-}\to He^+}$ 					& (3) & -\\
  5 & 2cx-He$^{0}$ 	& $\mathrm{He^{2+}+He^0\to He^0 + He^{2+}}$ 		& (1) & ++\\
  \hline
  \multicolumn{5}{c}{$\mathrm{He^{+}\to He^{2+}\;or\;He^0}$} \\
  \hline
  6 & ion-He$^{0}$ 	& $\mathrm{He^+ + He^0\to He^{2+} + He^0 + e^-}$ 	& (1) & +\\
  7 & cx$_2$-He$^{+}$ & $\mathrm{He^+ + He^+\to He^{2+} + He^{0}}$ 		& (1) & -\\
  8 & cx-He$^{2+}$ 	& $\mathrm{He^+ + He^{2+}\to He^{2+} + He^{+}}$ 	& (1) & \\
  9 & ion-H$^{0}$ 	& $\mathrm{He^+ + H^0\to He^{2+} + H^0 + e^-}$ 		& (1) & ++\\
  10 & cx-H$^{+}$ 	& $\mathrm{He^+ + H^+\to He^{2+} + H^0}$ 			& (1) & \\
  11 & ion-H$^{+}$ 	& $\mathrm{He^+ + H^+\to He^{2+} + H^+ + e^-}$ 		& (1) & \\
  12 & ion-e$^{-}$ 	& $\mathrm{He^+ + e^-\to He^{2+} + 2e^-}$ 			& (4) & \\
  13 & ph-ion 		& $\mathrm{He^+ + \gamma\to He^{2+} + e^-}$ 		& (6) & \\
  14 & cx-He$^{0}$ 	& $\mathrm{He^+ + He^0\to He^0 + He^+}$ 			& (1) & ++\\
  15 & cx$_0$-He$^{+}$ 	& $\mathrm{He^+ + He^+\to He^{0} + He^{2+}}$ 	& (1) & -\\
  16 & cx-H$^{0}$ 	& $\mathrm{He^+ + H^0\to He^0 + H^+}$ 				& (1) & ++\\
  17 & rc-e$^{-}$ 	& $\mathrm{He^+ + e^-\to He^{0}}$ 					& (5) & -\\
  \hline
  \multicolumn{5}{c}{$\mathrm{He^{0}\to He^{2+}\;or\;He^+}$} \\
  \hline
  18 & 2cx-He$^{2+}$ 	& $\mathrm{He^0 + He^{2+} \to He^{2+} + He^0}$ 	& (1) &-\\
  19 & ion-He$^{0}$ 	& $\mathrm{He^0 + He^0 \to He^+ + He^0 + e^-}$ 	& (1) & +\\
  20 & cx-He$^{+}$ 	& $\mathrm{He^0 + He^+ \to He^+ + He^0}$ 			& (1) & ++\\
  21 & ion-He$^{+}$ 	& $\mathrm{He^0 + He^+ \to He^+ + He^+ + e^-}$ 	& (7) & -\\
  22 & cx-He$^{2+}$ 	& $\mathrm{He^0 + He^{2+} \to He^+ + He^+}$ 	& (1) & -\\
  23 & 2cx-He$^{2+}$ 	& $\mathrm{He^0 + He^{2+} \to He^{2+} + He^0}$ 	& (1) & +\\
  24 & ion-He$^{2+}$ 	& $\mathrm{He^0 + He^0 \to He^+ + He^0 + e^-}$ 	& (1) & -\\
  25 & ion-H$^{0}$ 	& $\mathrm{He^0 + H^0 \to He^+ + H^0 + e^-}$ 		& (8) & ++\\
  26 & cx-H$^{+}$ 	& $\mathrm{He^0 + H^+ \to He^+ + H^0}$ 				& (1) & \\
  27 & ion-H$^{+}$ 	& $\mathrm{He^0 + H^+ \to He^+ + H^+ + e^-}$ 		& (1) & \\
  28 & ion-e$^{-}$ 	& $\mathrm{He^0 + e^- \to He^+ + 2e^-}$ 			& (4) & ++\\
  29 & ph-ion 	& $\mathrm{He^0 + \gamma \to He^+ + e^-}$ 				& (6) & \mathrm{(see\ the\ text)}\\
  \enddata
  \tablerefs{(1) \citet{barnett_1990}, (2) \citet{liu_2003a, havener_2005, panov_2002, minami_2008}, (3) \citet{arnaud_1985}, (4) \citet{janev_1987}, (5) \citet{aldrovandi_1973}, (6) \citet{bzowski_2013b, sokol_2014a}, (7) \citet{shevelko_2009}, (8) \citet{dubois_1989}.}
  \tablenotetext{a}{cx -- charge-exchange, 2cx -- double charge-exchange, ion -- ionization, rc -- recombination, ph-ion -- photoionization by solar EUV photons}
  \tablenotetext{b}{$++$ -- reaction dominates within at least a part of the energy range,\\ $+$ -- reaction contribute at least 10\% of the total reaction rate,\\ $-$ -- the mean free path against the reaction is above $10^6$~au}
\end{deluxetable*}

The primary source of information on the cross sections is Volume 1 of the ORNL CFADC ``Redbooks'' \citep{barnett_1990}. It is the same source as used by \citet{scherer_2014a}. However, for some reactions other sources are also used, as given in the references in Table~\ref{tab:appendixreaction}. Some of the reactions in Table~\ref{tab:appendixreaction} are repeated because the role of the reagents are different. Namely, in all reactions the first reagent and the first product is an ion or atom of interest, whereas the second is the component of the background plasma. Regardless of the further discussion, all of the reactions presented in the table are used in the numerical code.

The reactions in the table do not cover all of the possible reactions modifying the charge state of helium. We restrict possible reactions to the charge exchange, single ionization, and recombinations. The combinations of charge-exchange and ionization processes are neglected. For example, the reaction He$^{2+}$ + He$^0$ $\to$ He$^+$ + He$^{2+}$ + e$^{-}$ is neglected. In this reaction, the neutral atom simultaneously loses two electrons and one of them is captured by He$^{2+}$ while the other one is emitted freely. We also neglect double ionization processes, in which a helium atom loses two electrons due to impact with the other reagent. The cross sections for double ionization are typically more than 10 times smaller than for single ionization \citep{dubois_1988}. 

Another kind of reaction is elastic collisions. These collisions lead to the scattering of the particles, effectively decreasing the speed of the energetic ions or atoms. However, this process is highly ineffective for the conditions in the inner heliosheath. Even though the cross sections for elastic scattering can be higher than those for charge-exchange processes, scattering angles are typically small and decrease with increasing energy \citep{schultz_2008}. Consequently, significant change of energy requires many collisions. \citet{kubiak_2014} showed that helium atoms with speeds of $\sim$20~$\mathrm{km\,s^{-1}}$ thermalize at distances larger than 20\,000 au. Typical interaction speeds relevant for our study are much larger, and the typical distances of the heliospheric boundaries are shorter, thus the elastic collisions do not need to be included into the analysis.

\subsection{He$^{2+}$ to He$^+$ or He$^0$}\label{sec:he2}
A doubly ionized helium ion He$^{2+}$ can only be decharged. The second reagent needs to have at least one electron, thus it can be H$^0$, He$^0$, He$^+$, or e$^-$. Three types of reactions are possible: single charge exchange with H$^0$, He$^0$, He$^+$, double charge exchange with He$^0$, and recombination with e$^-$.

\begin{figure}
  \epsscale{.6}

  \plotone{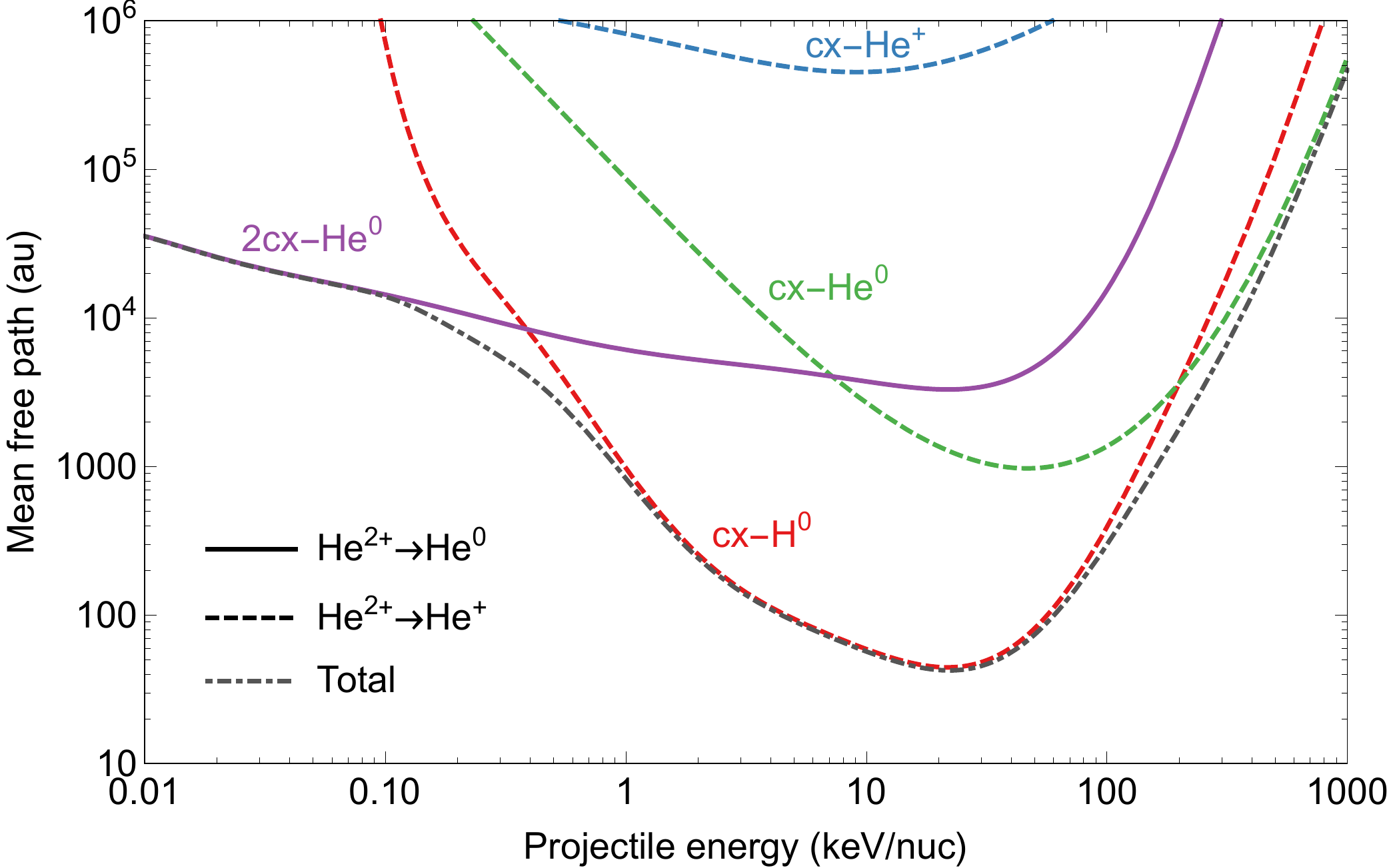}

  \plotone{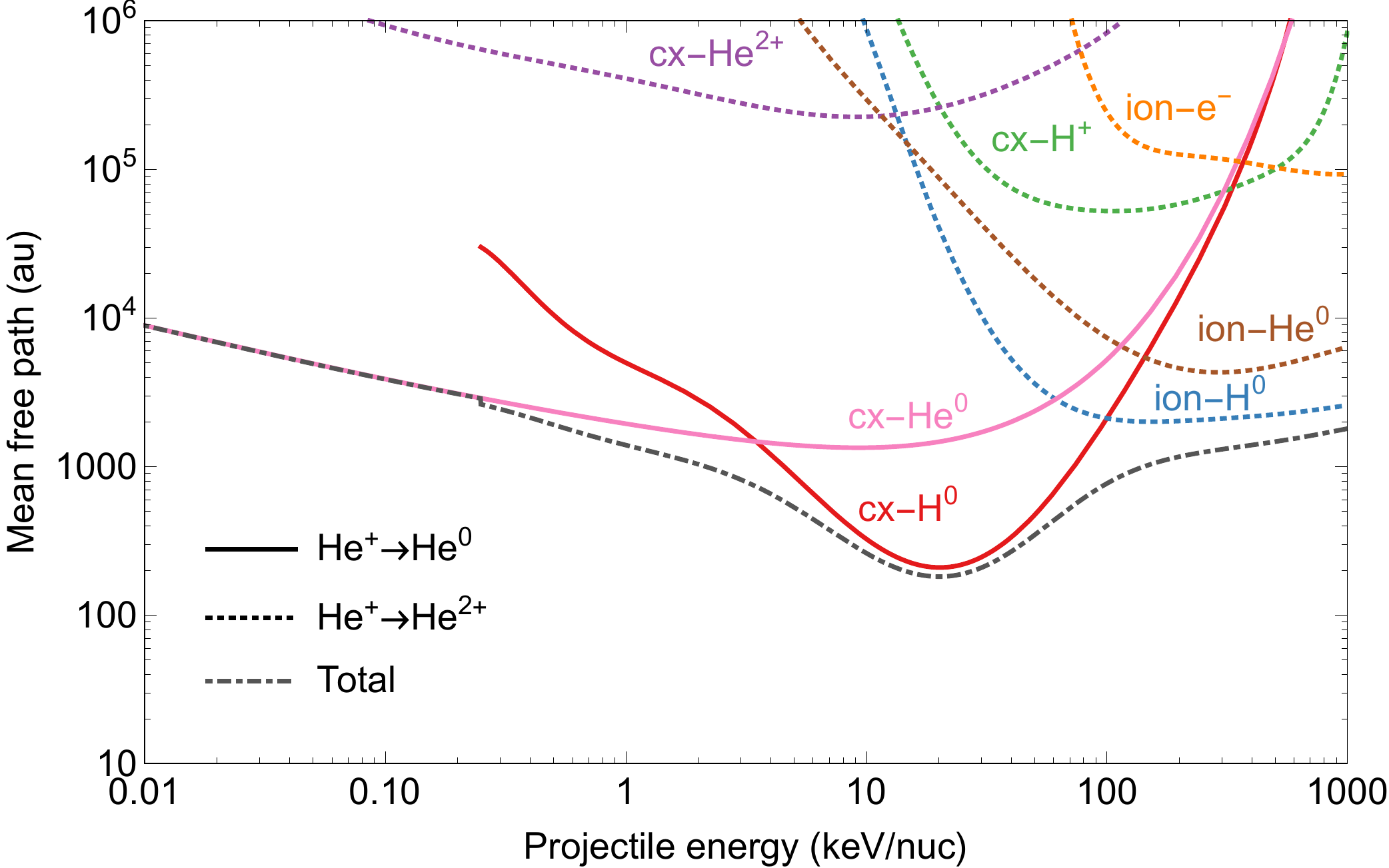}

  \plotone{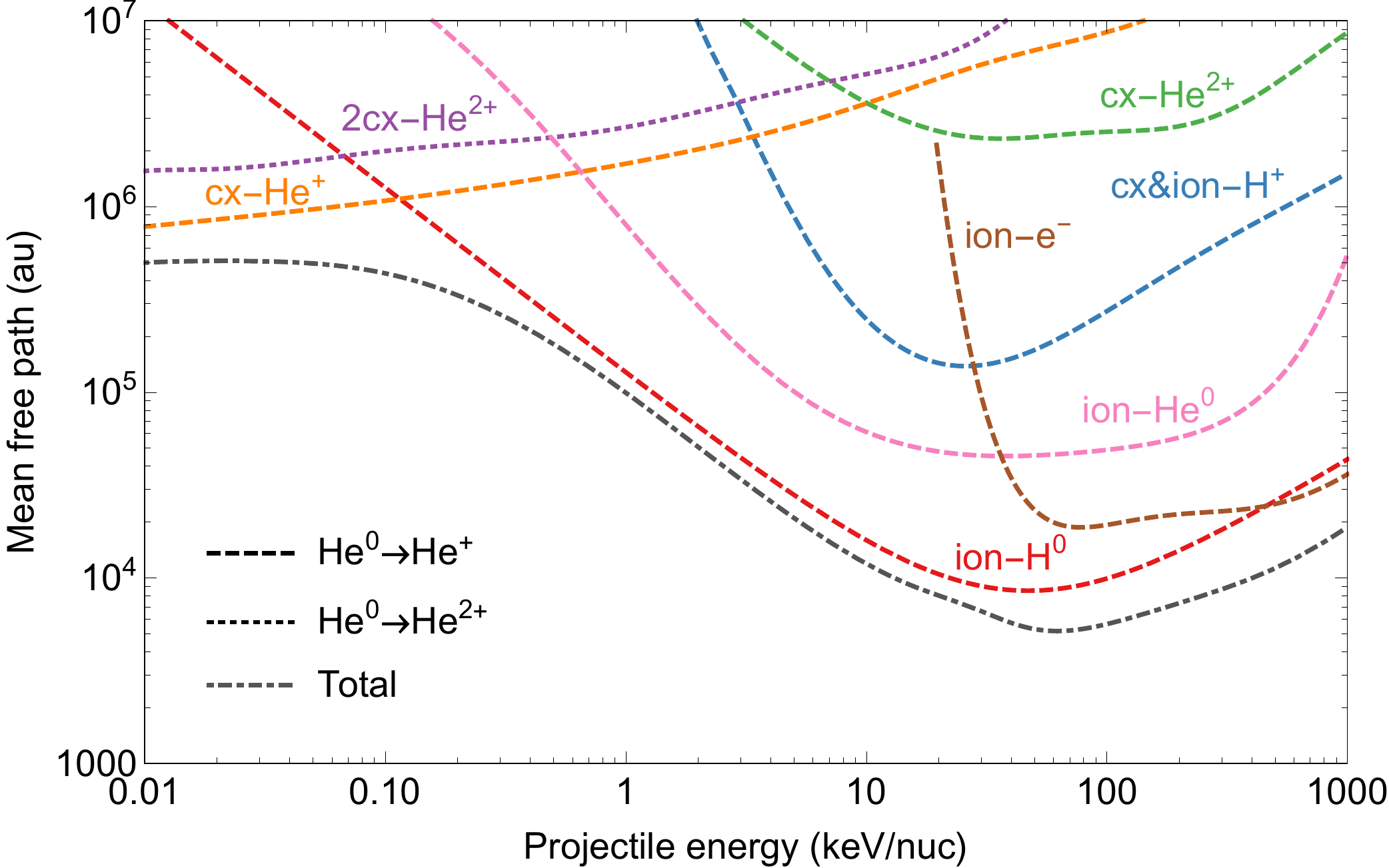}

  \caption{Mean free paths against the charge-change reactions of helium ions and atoms. Top to bottom panels present reactions for He$^{2+}$, He$^+$, and He$^{0}$. Solid, dashed, and dotted lines represent the final charge state of helium: He$^{0}$, He$^+$, and He$^{2+}$, respectively. The gray dash-dotted lines present the combined mean free path against all of the reactions.\label{fig:app:mfp}}
\end{figure}

The top panel of Figure~\ref{fig:app:mfp} shows the mean free paths against these reactions. As evident in this figure, important are the single and double charge exchange with He$^{0}$, as well as the charge exchange with H$^{0}$. The recombination with electrons is negligible. 

\subsection{He$^{+}$ to He$^{2+}$ or He$^0$}\label{sec:he1}
All components of the space plasma can change the charge state of He$^+$. Here, eight possible reactions for the ionization and four reactions for the neutralization are considered. The middle panel of Figure~\ref{fig:app:mfp} presents the mean free path against these reactions. For energies $\lesssim$100~keV/nuc more probable are neutralization processes, dominated by the charge exchange with He$^{0}$ and H$^{0}$. For higher energies, ionization is more probable. This is the reason for the steepening of the spectra of He ENAs for the highest considered energies. The recombination with electrons is also negligible. The ionization by electron impact contributes only little to the overall ionization processes, dominated by collisions with the neutrals. 

Another possible process is the photoionization. This process depends on the square of the distance from the Sun and does not depend on the energy of He$^+$, thus it is not plotted in the figure. The photoionization is negligible in the inner heliosheath, because the mean photoionization rate of He$^+$ at 1~au is $\sim$8$\times10^{-9}$~s$^{-1}$ \citep{sokol_2014a}, so the corresponding mean free path at 100~au is $1.2\times10^{6}$~au. 

The total mean free path of He$^+$ in the inner heliosheath is long, and exceeds 1000~au for 1 keV/nuc. This is the reason for the large tail-to-nose intensities ratio of He ENAs (see Section~\ref{sec:tail}). This ratio decreases for energies of a few tens of keV/nuc, for which the mean free path is the shortest. At those energies, helium ions do not penetrate so deeply the inner heliosheath.  

\subsection{He$^{0}$ to He$^{2+}$ or He$^+$}\label{sec:he0}
The reactions for ionization of He ENAs are treated only as losses for the observed fluxes and they are not included as a source term for new PUIs in the inner heliosheath. The bottom panel of Figure~\ref{fig:app:mfp} presents the mean free path of neutral helium against different ionization processes. The dominating process is the ionization by collisions with hydrogen. Charge exchange with helium ions has a large contribution only for the largest energies. The combined mean free path is so long that the inner heliosheath can be treated as completely transparent for He ENAs. 

Photoionization is also negligible in the inner heliosheath. The mean photoionization rate at 1~au of $10^{-7}$~s$^{-1}$ \citep{sokol_2014a} corresponds to the mean free path of 1 keV/nuc He ENAs of $4\times10^6$~au at 100~au from the Sun. However, this photoionization is an important contribution to the losses in the supersonic solar wind. For the lower energies, He ENAs can be photoionized close to the Sun.

\section{Details of numerical calculations}\label{appendix:numeric}
Here, the numerical scheme used to calculate the evolution of the distribution functions of helium ions in the inner heliosheath and then the resulting He ENA production is presented. 

\subsection{Flow lines}
The flow lines for a given potential can be obtained by integrating the plasma bulk velocity given by Equation~\eqref{eq:pot2vel}. The potential describes the plasma flow only in the inner heliosheath, thus the flow lines are calculated starting at the termination shock. The flow lines can be identified by an angular distance to the upwind direction $\psi$ due to rotational symmetry around the axis determined by this direction. The boundary conditions are $x(t=0)=r_\mathrm{s}\sin\psi$, $z(t=0)=r_\mathrm{s}\cos\psi$, where $r_\mathrm{s}$ is the distance to the termination shock. For both potentials we used, the flow lines are calculated for 180 different values of $\psi$ from $1\degr$ to $180\degr$ with a $1\degr$ step. Additionally, a flow line for $\psi=0.3\degr$ is calculated to provide a better coverage of the inner heliosheath close to the upwind direction.

Calculations of the evolution of the distribution functions along the flow lines require determination of different relative speeds and the corresponding cross sections of multiple reactions for all speed bins. A reasonable calculation time and a good accuracy can be approached with a non-uniform discretization of the flow lines. In this analysis, points along the flow lines are selected for time ticks $t_{n}=0.08n^{1.46}$~yr of the plasma bulk flow from the termination shock. The points are set for all integer values of $n$ up to some upper limit determined as described below. This time grid allows us to obtain a reasonable resolution close to the termination shock, where the changes of the plasma bulk flow are the largest (especially in the upwind hemisphere). Simultaneously, it allows for covering the far tail, which is important in the analysis of helium, with a reasonable total number of steps. The upper limit of $n$ is selected so that it is large enough for the last ticks on the flow line for $\psi=0.3\degr$ to be located closer than $1\degr$ to the downwind direction. Equivalently, it means that for all lines of sight pointing within $\psi<179\degr$, they cross this flow line. To achieve these requirements, the upper limit for $n$ was selected as 360 or 250 for the potentials $\Phi_1$ and $\Phi_2$, respectively. Moreover, this means that the model covers the distant tail up to $13000$~au and $7000$~au in the downwind direction for potentials $\Phi_1$ and $\Phi_2$, respectively. At each time tick, the following parameters of the plasma are tabulated: the position, the bulk flow velocity, the time, and the distance along the flow line from the termination shock. 

\subsection{Evolution of the distribution functions}
The spectrum at the termination shock (Equation~\eqref{eq:tsspectrum}) depends on two angles: the angular distance to the upwind direction ($\psi$) and the heliographic latitude ($\theta$). The first angle $\psi$ is discretized as described above and the second one $\theta$ with a $5\degr$ resolution: $0\degr$, $5\degr$, ..., $90\degr$. As described in Section~\ref{sec:spectra}, in this analysis it is assumed that the solar wind is symmetric with respect to the solar equator, thus only non-negative values of $\theta$ need to be calculated. The circle determined by a certain value of angle $\psi$ often does not cover all heliographic latitudes, and thus only a minimal set covering the actual range is calculated. For example, for $\psi=37\degr$, the range of the heliographic latitudes is $(-31\fdg9, 42\fdg1)$, thus we calculate the evolution of helium ions for $\theta=0\degr,\,5\degr,\,...,45\degr$. These two angles yield 2078 possible combinations in total. In each of these cases the evolution is calculated separately.

The assumed spectrum of helium ions downstream of the termination shock given by Equation~\eqref{eq:tsspectrum} is tabulated at the termination shock to 1000 speed bins with the bin width of 20~$\mathrm{km\,s^{-1}}$. At the termination shock, the distribution function in each speed bin is evaluated at the center speed, i.e., at 10, 30, 50, ..., 19990~$\mathrm{km\,s^{-1}}$. They are assumed isotropic in the plasma frame throughout the heliosheath, consequently, the mean relative speed between helium ions and the other substrate, averaged over one of the speed bins, can be approximated as a square root of a sum of squares of the bulk speed, the bin center speed, and the characteristic thermal speed of the other substrate. This grid allows to obtain a reasonable accuracy of calculation of the relative speeds. A typical speed of the bulk flow ($\sim$100~$\mathrm{km\,s^{-1}}$) combined with the bin center speeds results in a difference between the edges of each bin compared to the center speed less than $\sim$5\%. This covers the energy range from 0.5 eV/nuc to 2 MeV/nuc. 

The time evolution of the distribution functions of helium ions is given by Equations~\eqref{eq:simtransport}--\eqref{eq:collterm2}. We solve these equations numerically using the evolution from the time $t_{n}$ to $t_{n+1}$ given as:
\begin{align}
 f_{\mathrm{He^+},{n+1}}&=
  e^{-\left(\alpha_{1\to0}+\alpha_{1\to2}\right)\left(t_{n+1}-t_n\right)}f_{\mathrm{He^+},{n}}
 +\left(1-e^{-\alpha_{2\to1}\left(t_{n+1}-t_n\right)}\right)f_{\mathrm{He^{2+}},{n}}
 +(t_{n+1}-t_n)\frac{v}{3}\frac{\partial f_{\mathrm{He^+},{n}}}{\partial v}\left(\vec{\nabla}\cdot\vec{u}\right), \label{eq:evhe1}\\
 f_{\mathrm{He^{2+}},{n+1}}&=
  e^{-\left(\alpha_{2\to0}+\alpha_{2\to1}\right)\left(t_{n+1}-t_n\right)}f_{\mathrm{He^{2+}},{n}}
 +\left(1-e^{-\alpha_{1\to2}\left(t_{n+1}-t_n\right)}\right)f_{\mathrm{He^{+}},{n}}
 +(t_{n+1}-t_n)\frac{v}{3}\frac{\partial f_{\mathrm{He^{2+}},{n}}}{\partial v}\left(\vec{\nabla}\cdot\vec{u}\right),\label{eq:evhe2}
\end{align}
In these equations, $s_{n+1}$ and $s_n$ denote the distances along the flow line at the times $t_{n+1}$ and $t_n$, respectively. The bulk plasma speed $u=|\vec{u}|$ is calculated as the mean value between the plasma speed at two time points. The divergence of the plasma flow $\vec{\nabla}\cdot\vec{u}$ is 0 for potential $\Phi_1$, and constant, equal $-\alpha_\mathrm{cx}/(2\gamma)$ for potential $\Phi_2$ (see Section~\ref{sec:flow}). The reaction rates $\alpha_{i\to j}$ are calculated for the mean plasma conditions between these two data points (see Appendix~\ref{appendix:crosssec}). The differentiations of the distribution functions with respect to the atom speed are calculated numerically using central differences, except for the first and last speed bin. The formulas used for calculations of the cross sections can be used only within certain limited ranges of the relative velocities, for which their provide proper values. The formulas cover the ranges at which the contribution of each reaction is important. However, these limitations lead to non-smooth transitions at the edges of these ranges. Consequently the calculation of the central differences can be affected. To avoid this problem, logarithms of the distribution functions obtained from Equations \eqref{eq:evhe1} and \eqref{eq:evhe2} are Gaussian-smoothed. The width of the Gauss function used as a weights is growing linearly from 1 bin at the lowest bin to 25 bins at the highest.

With the presented scheme, the resulting distribution function in each time bin is tabulated to the same speed bins as described above. 

\subsection{Integration of the ENA signal and survival probabilities}
The considered flows of the plasma in the inner heliosheath have rotational symmetry, and each flow line is confined to the plane determined by the upwind direction and the direction in which the flow line starts at the termination shock. This allows us to simplify the integration of the ENA signal according to Equation~\eqref{eq:enaintensityv}. 

First, the plane containing the upwind direction and the line of sight in which the ENA flux is calculated is determined. In this plane, the heliolatitudes of the beginnings of the flow lines at the termination shock are determined for each of the available angular distances from the upwind direction. The distributions of helium ions along these flow lines are determined using linear interpolation between the calculated distributions on the $5\degr$-resolution grid. From these ion distributions we calculate the ENA signal for the selected line of sight.

The distribution functions of helium ions are tabulated in a finite number of grid points spread over the flow lines, which are not aligned with the lines of sight in which the ENA signal is integrated. A set of pairs of grid points is chosen for each line of sight. The grid points in each pair are selected so that the section between them intersects the line of sight and that these points are close to the line of sight (within $2\degr$). The distribution functions are linearly interpolated to those intersections. Additionally, the pairs are selected so that the distance between the consecutive intersections is larger than 1 au and smaller than 3 au. The integral of the ENA signal is approximated by a numerical quadrature using the trapezoidal rule applied to the distribution functions at these intersections. 

The intensities of He ENAs calculated from Equation~\eqref{eq:enaintensityv} represent the spectra in the Sun's frame. The plasma is flowing outwards from the Sun, thus the speed in the Sun's frame $v$ is smaller than the speed in the plasma frame $v'$ (Equation~\eqref{eq:vconversion}). The differential intensity is calculated for the same speed bins in the Sun's frame as the distribution functions, and the distribution functions are interpolated to the speed in the plasma frame. The spectra presented in this paper are plotted using the third order spline interpolation between the energy bins resulting from the used speed grid. The accuracy of this interpolation is within $\pm$2\%, assessed from the finite differences of the intensity spectra. As a result, some high-speed bins in the plasma frame are out of the range, but these speeds are out of the scope of this analysis and the distribution function there is assumed zero. 

The integral inside the exponent of the survival probabilities $w_\mathrm{sur}(r,v)$ (Equation~\eqref{eq:survprob}) is calculated separately inside the termination shock and in the inner heliosheath. Inside the termination shock we include only the contribution from the photoionization, and the integral is equal to $\nu_\mathrm{He,0}r_0(1-r_0/r_\mathrm{s})/v$, where $\nu_\mathrm{He,0}$ is the photoionization at $r_0=1\,\mathrm{au}$, and $r_\mathrm{s}=90\,\mathrm{au}$ is the distance to the termination shock. The contribution of the other processes in the supersonic solar wind is negligible \citep{bzowski_2013b}. In the inner heliosheath, the integral is calculated based on the plasma parameters obtained on the same grid points in which the distribution functions are obtained. 

\bibliographystyle{aasjournal}
\bibliography{library}


\end{document}